%% file: main.tex
\documentclass[10pt,twocolumn,letterpaper]{article}

\usepackage{multirow}

\usepackage[pagenumbers]{iccv} % To force page numbers, e.g. for an arXiv version


\definecolor{iccvblue}{rgb}{0.21,0.49,0.74}
\usepackage[pagebackref,breaklinks,colorlinks,allcolors=iccvblue]{hyperref}

\newcommand{\myparagraph}[1]{\smallskip\noindent\textbf{#1}}

\def\paperID{276} % *** Enter the Paper ID here
\def\confName{ICCV}
\def\confYear{2025}

\title{BrainMRDiff: A Diffusion Model for Anatomically Consistent\\ Brain MRI Synthesis}
\author{Moinak Bhattacharya\\
Biomedical Informatics\\
Stony Brook University, US\\
\and
Saumya Gupta\\
Computer Science\\
Stony Brook University, US\\
\and
Annie Singh\\
Ram Manohar Lohia Hospital, \\India\\
\and
Chao Chen\\
Biomedical Informatics\\
Stony Brook University, US\\
\and
Gagandeep Singh\\
Radiology\\
Columbia University, US\\
\and
Prateek Prasanna\\
Biomedical Informatics\\
Stony Brook University, US\\
}

\begin{document}
\maketitle
\input{sec/0_abstract}    
\input{sections/1_introduction}

\input{sections/2_related_works}
\input{sections/3_method}
\input{sections/4_results}
\input{sections/5_conclusion}
% \\
% \\
% \textbf{Acknowledgments}
% The reported research was partly supported by NIH 1R21CA258493-01A1, NIH 75N92020D00021
% (subcontract), and the OVPR and IEDM seed grants at Stony Brook University. The con-
% tent is solely the respon- sibility of the authors and does not necessarily represent the official views of the National Institutes of Health.
    % \clearpage
{
    \small
    \bibliographystyle{ieeenat_fullname}
    \bibliography{main}
}
\input{supplementary}


\end{document}

%% file: sec/0_abstract.tex
\begin{abstract}
% The ABSTRACT is to be in fully justified italicized text, at the top of the left-hand column, below the author and affiliation information.
% Use the word ``Abstract'' as the title, in 12-point Times, boldface type, centered relative to the column, initially capitalized.
% The abstract is to be in 10-point, single-spaced type.
% Leave two blank lines after the Abstract, then begin the main text.
% Look at previous \confName abstracts to get a feel for style and length.
Accurate brain tumor diagnosis relies on the assessment of multiple Magnetic Resonance Imaging (MRI) sequences. However, in clinical practice, the acquisition of certain sequences may be affected by factors like motion artifacts or contrast agent contraindications, leading to suboptimal outcome, such as poor image quality. This can then affect image interpretation by radiologists. Synthesizing high quality MRI sequences has thus become a critical research focus. Though recent advancements in controllable generative AI have facilitated the synthesis of diagnostic quality MRI, ensuring anatomical accuracy remains a significant challenge. Preserving critical structural relationships between different anatomical regions is essential, as even minor structural or \textit{topological} inconsistencies can compromise diagnostic validity. In this work, we propose BrainMRDiff, a novel topology-preserving, anatomy-guided diffusion model for synthesizing brain MRI, leveraging brain and tumor anatomies as conditioning inputs. To achieve this, we introduce two key modules: Tumor+Structure Aggregation (TSA) and Topology-Guided Anatomy Preservation (TGAP). TSA integrates diverse anatomical structures with tumor information, forming a comprehensive conditioning mechanism for the diffusion process. TGAP enforces topological consistency during reverse denoising diffusion process; both these modules ensure that the generated image respects anatomical integrity.  Experimental results demonstrate that BrainMRDiff surpasses existing baselines, achieving performance improvements of 23.33\% on the BraTS-AG dataset and 33.33\% on the BraTS-Met dataset. Code will be made publicly available soon. %upon acceptance.
\end{abstract}

%% file: sections/1_introduction.tex
\section{Introduction}
Each year, an estimated 300,000 individuals worldwide are diagnosed with brain tumors~\cite{bray2024global}. Multiparametric Magnetic Resonance Imaging (MRI) serves as the gold standard for detecting and characterizing these tumors, providing high-resolution information necessary for accurate diagnosis, treatment planning, and monitoring. However, in clinical practice, the acquisition of MRI sequences faces significant limitations. Patients may be unable to remain still for extended periods, may have contraindications to contrast agents, or facilities may have hardware constraints—-- often resulting in incomplete imaging sequences that compromise diagnostic accuracy and treatment planning. 

\begin{figure}[t]
    \centering
\includegraphics[width=1\linewidth]{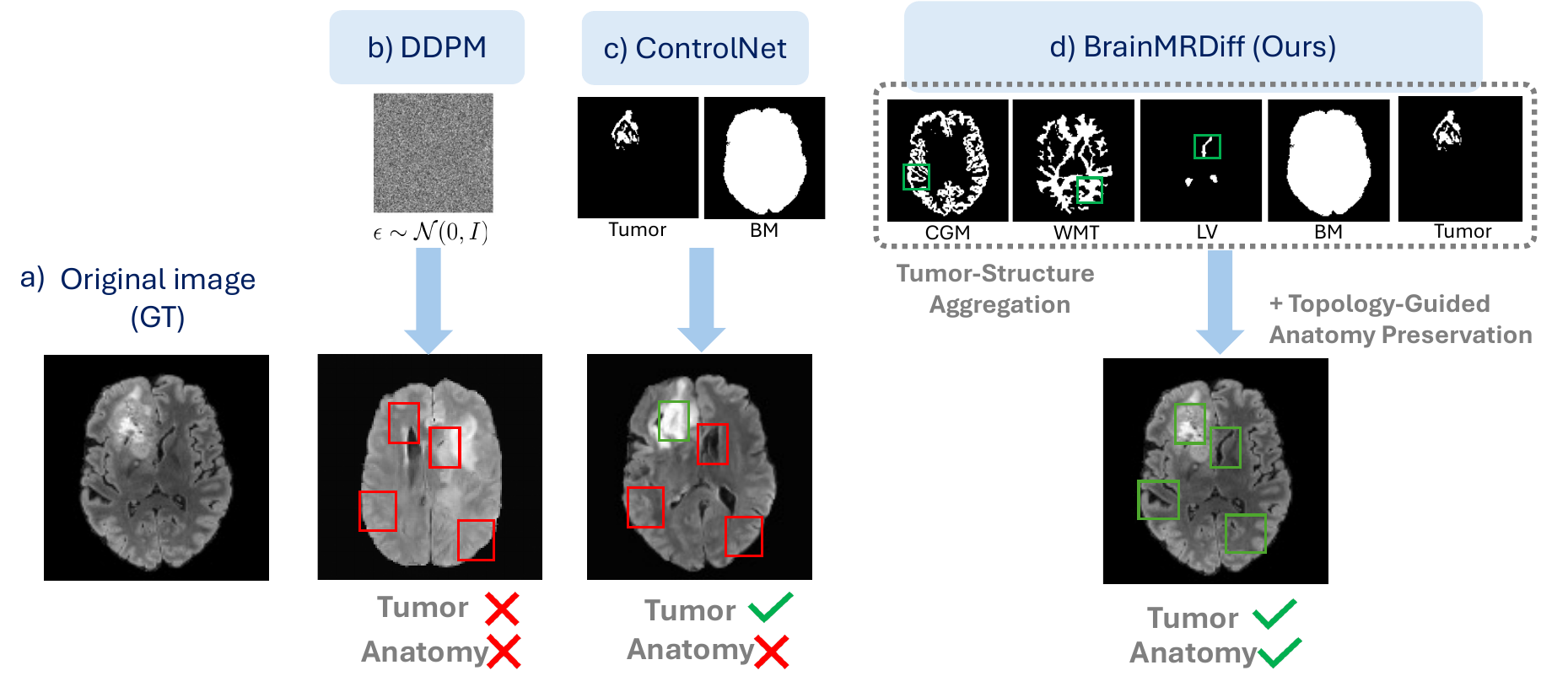}
    \caption{
    \textbf{Overview of our proposed work.} Baseline methods exhibit limitations in generating MR images with faithful anatomical representations. In contrast, our proposed BrainMRDiff framework integrates anatomical constraints—specifically WMT, CGM, LV, and tumor masks as control inputs—to produce MR images that accurately reflect anatomical structures}
    %\saum{would a layout like this be better for the teaser?} \pp{yes} \gs{yes- this looks good- need to add GT image}
    
    \label{fig:teaser}
\end{figure}

The advent of deep learning models, and in particular, generative models, has revolutionized this field. They have been explored as a means to synthesize missing or low-quality MRI sequences. For instance, Generative Adversarial Networks~\cite{goodfellow2020generative} have been proposed to generate high-quality, realistic volumetric sequences for brain imaging~\cite{alex2017generative,kwon2019generation,hu2021bidirectional}. Recently, diffusion models~\cite{ho2020denoising,rombach2022high}, have garnered increasing attention in the field~\cite{pinaya2022brain} due to their superior image generation quality and training stability. An important extension of diffusion models is the introduction of conditioning mechanisms, which allow image generation to be guided by specific input conditions, such as textual descriptions~\cite{saharia2022photorealistic,ruiz2023dreambooth}, pose information~\cite{tseng2023consistent}, or segmentation maps~\cite{zhang2023adding}. These conditional diffusion models have been explored for medical image generation, leveraging conditions such as segmentation masks~\cite{konz2024anatomically}, radiomics filters~\cite{bhattacharya2024radgazegen}, gaze patterns~\cite{bhattacharya2024gazediff}, and topological constraints~\cite{xu2024topocellgen}. While these conditioning strategies have significantly improved the fidelity of generated medical images, they often fail to explicitly account for anatomical structures during the generation process, limiting their applicability in real-world clinical settings. (see \cref{fig:teaser})
%\saum{Refer to the teaser here (teaser can show poor anatomical results of baselines compared to ours).}

In this work, we aim to address the critical challenge of ensuring the accuracy of anatomical regions when generating brain MRI sequences. Our key innovation lies in leveraging anatomical knowledge to guide the synthesis of brain MRI sequences in order to preserve critical structural features. Existing generative models, while effective at producing visually convincing images, often fail to maintain the intricate structural relationships between brain regions and tumors that are essential for clinical interpretation. In particular, brain tumors exhibit highly heterogeneous morphology, making their accurate synthesis especially difficult. Without explicit constraints, generative models tend to introduce distortions or unrealistic structures, reducing the reliability of synthesized MRI sequences for diagnosis and treatment planning. To address this, we propose leveraging anatomical priors to guide the generation process, ensuring that key brain structures and tumors maintain their natural topology throughout synthesis. %By incorporating anatomical guidance through multiple region-specific masks and enforcing topological consistency, particularly in heterogeneous tumor regions, 
Through this, we can generate synthetic MRI sequences that not only look realistic but also retain the precise structural characteristics necessary for accurate diagnosis. This anatomy-aware approach represents a fundamental shift from purely appearance-based synthesis to structure-preserving generation that aligns with clinical requirements. 

We propose BrainMRDiff, a topology-guided diffusion model designed to preserve the structural details of anatomical regions in synthesized MRI sequences. Our approach conditions the generation process on multiple anatomical masks, including White Matter Tracts (WMT), Cortical Gray Matter (CGM), Lateral Ventricles (LV), Brain Masks (BM), and tumor segmentation masks. Notably, no existing work has investigated the combined utilization of these anatomical structures as conditioning inputs for diffusion-based image synthesis. However, a key challenge remains--—the generated anatomical structures must maintain their topological consistency, particularly in tumor regions, where morphology varies significantly across patients. To tackle this variability, we introduce a topology-preserving loss function that enforces structural fidelity in tumor regions. 
% To the best of our knowledge, we are the first to integrate topological constraints within diffusion models for brain MRI synthesis. 
Our approach bridges a critical gap in medical image generation, ensuring that synthetic MRI scans not only appear realistic but also retain anatomical accuracy, making them more clinically applicable.

To summarize, our contributions are as follows:
%\textbf{Contributions} The primary contributions of this work are summarized as follows:
\begin{itemize}
    \item We introduce BrainMRDiff, a novel structure-aware and topology-preserving diffusion model that incorporates anatomical structures as conditioning inputs while ensuring the topological integrity of tumor structures.
    \item We propose a Tumor-Structure Aggregation Module, which integrates multiple anatomical structures along with tumor morphology into a unified control mechanism. This aggregated representation serves as a conditional guidance input to the diffusion model, facilitating the generation of anatomically coherent brain MRI sequences.
    \item To enforce topological consistency, we develop a Structure-Aware Topology Module, which computes the persistence diagram (PD) from the masked tumor region within the predicted noise of the diffusion model. The topological loss is then derived by minimizing the distance between the ground-truth and predicted PDs, ensuring structural fidelity in the generated tumor regions.
\end{itemize}

To the best of our knowledge, this is the first \textit{anatomy-aware and topology-guided diffusion model} designed for brain MRI generation. Our approach uniquely integrates multiple anatomical structures as conditioning inputs while explicitly preserving tumor topology through PD computations on the predicted noise within the diffusion model.

%% file: sections/2_related_works.tex
\section{Related Works}
\begin{figure*}[t]
    \centering
    \includegraphics[width=0.8\linewidth]{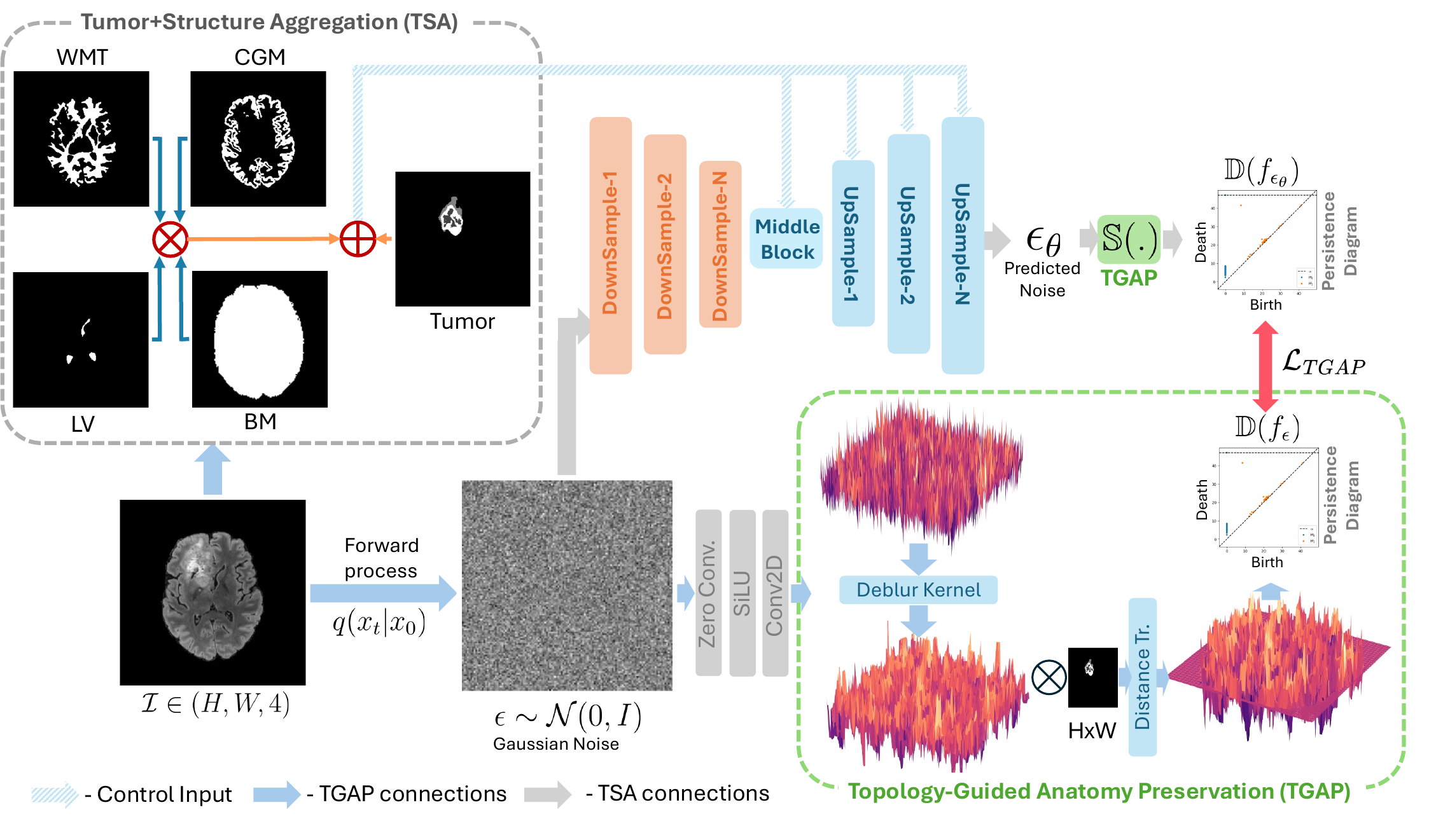}
    \caption{\textbf{BrainMRDiff architecture.} Our proposed method consists of two components: a) Tumor+Structure Aggregation (TSA) module which aggregates the different anatomical structures and tumor segmentation masks as a unified control to the diffusion model, b) Topology-Guided Anatomy Preservation (TGAP) module which enforces topological constraints  to ensure high fidelity tumor region generation.}
    %\pp{incomplete}  \saum{what does dashed arrow (in top row) vs solid arrow mean? And gray vs blue arrow? Include a legend box if needed} \saum{Try to use some english text too: like `Predicted noise $\epsilon_\theta$' instead of just $\epsilon_\theta$; Persistence diagram D instead of just D etc} \saum{I think $\epsilon \approx N$ is wrong because that is during inference and not training} \saum{what is MB?} \saum{write the shortform TSA and TGAP in brackets beside the module name} \saum{In the TSA, there is no outgoing arrow from the $\oplus$ operator. Maybe put tumor to the right of WMT/CGM/etc so that the outgoing arrow can be connected to the dashed arrow}
    
    \label{fig:architecture}
\end{figure*}

\myparagraph{Medical diffusion models.} 
%\saum{I combined the medical diffusion models and controllable diffusion models into one subsection. I shortened it all but retained the citations.}
%\pp{There is nothing in the paragraph about medical DMs}
Diffusion Models~\cite{ho2020denoising} have transformed the field of image generation in recent years~\cite{song2020score,rombach2022high,saharia2022photorealistic,ruiz2023dreambooth,radford2021learning,mou2024t2i}. 
%However, the inception of diffusion-based generative models can be traced back to score-based generative modeling using stochastic differential equations (SDEs)~\cite{song2020score}. Latent Diffusion Models (LDMs) operate within the latent space of images to execute the diffusion process~\cite{rombach2022high}. Further modification lead to using text as conditions to the diffusion models for guided image generation~\cite{saharia2022photorealistic,ruiz2023dreambooth}. Text-to-image (T2I) diffusion models integrates the image latent space embeddings with the text-embeddings generated from CLIP~\cite{radford2021learning} and similar architectures. More improvements on T2I models were proposed when accurate controlling paradigms, for example color, structure, etc, were used from more granular image generation~\cite{mou2024t2i}. 
In the medical imaging domain, diffusion models have been applied to several tasks such as synthetic image generation~\cite{khader2023denoising,kazerouni2023diffusion,yoon2023sadm,kim2022diffusion,pinaya2023generative}, image enhancement~\cite{ma2023pre,kachouie2008anisotropic,selim2023latent,liu2023esdiff}, anomaly detection~\cite{behrendt2025guided, wolleb2022diffusion}, segmentation~\cite{rahman2023ambiguous,wu2024medsegdiff,wu2024medsegdiff,chen2023berdiff,dong2024diffusion,guo2023accelerating}, etc. Particularly in the case of brain MRI, diffusion models have been extensively applied to tasks such as motion correction in parallel MRI~\cite{chen2025joint}, super-resolution to enhance MRI image clarity~\cite{ma2025diffusion}, tumor segmentation~\cite{qin2025btsegdiff}, among others.
%analysis, demonstrating their effectiveness in synthesis, reconstruction, and anomaly detection. In a work, a self-calibrating score-based diffusion model has been proposed for motion correction in parallel MRI~\cite{chen2025joint}. In another work, a diffusion model-based MRI super-resolution synthesis method to enhance image clarity~\cite{ma2025diffusion}. In a work, diffusion models have been used for tumor segmentation~\cite{qin2025btsegdiff}. 
However, to generate clinically accurate medical images some conditioning ought to be used to guide the diffusion models.

\myparagraph{Controllable diffusion models.} In addition to text-to-image diffusion models, several other controls can be used to guide image generation~\cite{huang2023composer,zhang2023adding,mou2024t2i,li2023gligen,zhao2023uni,qin2023unicontrol,tseng2023consistent,shen2023advancing,ma2024follow,alexanderson2023listen,shen2023difftalk,lee2024scribble}. 
%This can be primarily achieved in two strategies. One approach is training a diffusion model from scratch using single or multiple controls~\cite{huang2023composer}. And, another approach is finetuning lightweight adapters~\cite{zhang2023adding,mou2024t2i,li2023gligen}. These methods mostly takes a single control as input to guide image generation. Several recent methods incorporates more than one controls for image generation~\cite{zhao2023uni,qin2023unicontrol}. In addition to image controls, other guidance mechanismcs such as pose~\cite{tseng2023consistent,shen2023advancing,ma2024follow}, audio~\cite{alexanderson2023listen,shen2023difftalk}, scribbles~\cite{lee2024scribble}, etc are also used as controls. 
%These methods are not just limited to image generation but also are able to generate videos~\cite{ho2022video,esser2023structure,chen2024videocrafter2,blattmann2023stable,blattmann2023align}. 
With these advancements, several methods have also been proposed in medical imaging domain. RoentGen~\cite{chambon2022roentgen} generates Chest X-ray (CXR) images from radiologists' reports. ControlPolypNet generate synthetic polyps from sizes and locations aas controls \cite{sharma2024controlpolypnet}.
GazeDiff~\cite{bhattacharya2024gazediff} and RadGazeGen~\cite{bhattacharya2024radgazegen} generate CXR images using radiologist's eye gaze patterns and radiomics features.
%GazeDiff~\cite{bhattacharya2024gazediff} proposes to generate CXR images using radiologist's eye gaze patterns as control for zero shot disease classification. RadGazeGen~\cite{bhattacharya2024radgazegen} combines radiomics filters and radiologist's eye gaze patterns as control for clinically accurate CXR image generation. 
Recently, a few methods have been proposed to use anatomy as a control to diffusion model~\cite{zhang2024anatomy,konz2024anatomically}. However, no existing methods have been proposed to aggregate different anatomical structures and use them as a unified control to condition the diffusion model. 
%\saum{you mentioned earlier about MAISI using 127 anatomical controls --- how are we different from them?}

\myparagraph{Topological data analysis.} Topological Data Analysis (TDA)~\cite{carlsson2009topology} is an adaptation of algebraic topology which has found its versatile application in machine learning domain. Several approaches have proposed using different concepts of TDA, such as persistent homology (PH)~\cite{edelsbrunner2002topological,edelsbrunner2010computational,peng2024phg,asaad2022persistent}, discrete morse theory (DMT)~\cite{dey2019road,hu2021topology,banerjee2020semantic,batzies2002discrete}, etc. Other recent methods such as topological interctions~\cite{gupta2022learning}, centerline transforms~\cite{shit2021cldice,shi2024centerline,wang2022ta}, etc. With the advent of diffusion models, several topology-aware diffusion models have been proposed~\cite{hu2024topology,song2024topo,zhu2024controltraj,parktopology}. have found its applicability in several medical imaging applications. Topology has been extensively used for cancer research~\cite{zhao2023single,singh2014topological,yadav2023histopathological,wang2025topotxr}. Topology has also been used for image registration and reconstruction~\cite{chu2023topology,sun2022topology}. Recently, topology-based diffusion models have made inroads into the domain where constraint guide the generation of the desired topology~\cite{gupta2024topodiffusionnet,xu2024topocellgen}. A few recent works have proposed topology-aware anatomy segmentation methods in medical imaging~\cite{berger2024topologically,yang2024benchmarking}. However, existing works \cite{gupta2024topodiffusionnet,xu2024topocellgen} impose topology constraints that either enforce single-pixel connected components \cite{xu2024topocellgen} or generate large components with poor boundary details \cite{gupta2024topodiffusionnet}. Both approaches are primarily designed for counting tasks rather than preserving object details. As a result, these methods fail to retain fine details, such as tumor appearance, which is the focus of our work.
%\pp{Another short related work paragraph on brain MRI generation work}

In summary, there are no existing methods that deal with multiple anatomical structures from the brain MRI sequences as control and preserve the topology of the tumor structure from the predicted noise of the diffusion model. To address these shortcomings, we propose aggregating different anatomical structures while preserving spatial heterogeneity by enforcing topological consistency in tumor regions.
%\pp{Incomplete}

%% file: sections/3_method.tex
\section{Method}
\cref{fig:architecture} provides an overview of our proposed work, BrainMRDiff. Our goal is to generate anatomically-accurate brain MRI scans.
%In this work, we propose to generate anatomically aware and topologically-guided brain MRI scans. 
BrainMRDiff consists of two key components: Tumor+Structure Aggregation (TSA) and Topology-guided Anatomy Preservation (TGAP).
%\pp{Saumya - I dont think this is technically correct. maybe Anatomy-aware topology?} \saum{Agreed. We can use `structure' and `topology' interchangeably but not together. Anatomy-aware Topology is good. We could also do Topology-guided Anatomy Preservation (TGAP) module.}. 
The TSA module combines anatomical structures such as the Brain Mask (BM), White Matter Tract (WMT), Cortical Gray Matter (CGM), and Lateral Ventricles (LV) with tumor masks %such that these act as
to provide conditional control to the diffusion model. These masks ensure spatial correctness and detail preservation of the anatomical structures in the generated image.
%Given that these are complex anatomical regions and arranged densely in close proximity to each other, it becomes important to ensure the accuracy in the structural arrangement. 
While these conditional controls significantly improve the generated image quality for BM, WMT, CGM, and LV, they are still not powerful enough to capture the heterogeneity of tumor regions. To capture irregular tumor patterns, we propose stronger constraints to guide the diffusion model. Specifically, we propose the TGAP module to enforce topological constraints. It does so by leveraging tools such as persistent homology~\cite{edelsbrunner2002topological,edelsbrunner2010computational} and diagrams from topological data analysis~\cite{carlsson2009topology}. Enforcing topological constraints ensures higher-fidelity tumor region generation.
%... For this, the TGAP module ensures the topological correctness of the tumor structure by computing a distance between the persistent diagrams ground-truth and predicted structures.
%\pp{Minak- you told me that this was only enformed on tumor. I am confused now. is it just tumor or all structures?}. 

The rest of the section is organized as follows. We begin with a brief discussion of the preliminaries of diffusion models. Next, we present the TSA module in~\cref{sec:tsa} and the TGAP module in~\cref{sec:tgap}. Finally, we integrate these modules into an end-to-end training paradigm, as described in~\cref{sec:training}.

%In the following subsections, we discuss in detail the different sub-components of the architecture.\\
\begin{figure}[t]
    \centering
    \includegraphics[width=1\linewidth]{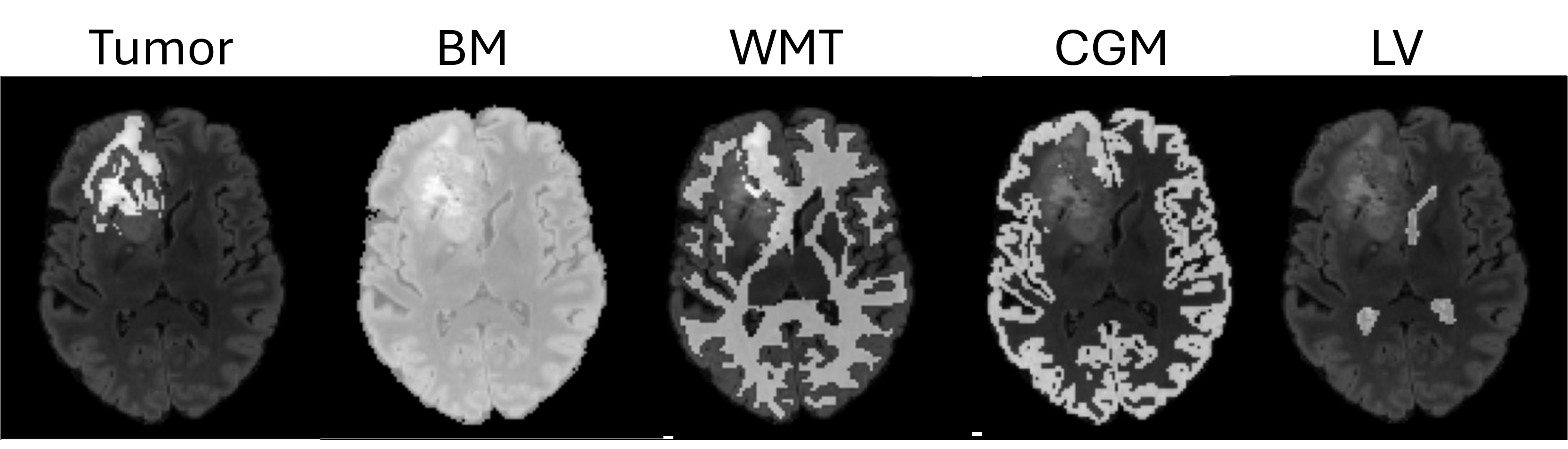}
    \caption{\textbf{Tumor and Anatomy Structures.} The tumor mask and the different anatomical structures namely whole Brain Mask (BM), White Matter Tracts (WMT), Cortical Gray Matter (CGM), Lateral Ventricles (LV) are shown overlaid on top of a FLAIR scan. 
    %\saum{increase text size inside figure}
    }
    \label{fig:structure}
\end{figure}

\myparagraph{Preliminaries.} %A diffusion model applies the concept of diffusion for image generation tasks~\cite{sohl2015deep}. 
% Initially, Gaussian Noise is injected to the image and then the diffusion model learns to predict the image from the noise. The noise N is added at different timesteps, and this process is known as the forward process. This noise is then fed to a UNet architecture which denoises xxx to predict the image xxx and this process is known as the reverse process. 
%Diffusion probabilistic models~\cite{ho2020denoising}, often referred to as 
Diffusion models~\cite{ho2020denoising} are generative models that are parameterized forms of Markov chains trained using variational inference. Consider an input distribution $\textbf{x}\sim q(\textbf{x}_0)$. If $T$ is the number of total time steps, the forward process $q(\textbf{x}_{1:T}|\textbf{x}_0)$ is defined as adding Gaussian noise to \textbf{x} using to a variance schedule $\beta_t\in\{\beta_0, \beta_1, ..., \beta_T\}$: %, where $T$ is the number of total time steps.  
% ...
\begin{equation}
    \begin{aligned}
        q(\textbf{x}_{1:T}|\textbf{x}_0) := \prod_{t=1}^T q(\textbf{x}_t|\textbf{x}_{t-1}); \\
        q(\textbf{x}_t|\textbf{x}_{t-1}) := \mathcal{N}(\textbf{x}_t; \sqrt{1-\beta_t}\textbf{x}_{t-1}, \beta \textbf{I})
    \end{aligned}
\end{equation}
For the reverse process, $\textbf{x}_{t-1}$ is recovered from $\textbf{x}_t$ by learning $p_\theta$ which aims to approximate the posterior distribution $q(\textbf{x}_{t-1}|\textbf{x}_t, \textbf{x}_0)$:
\begin{equation}
    p_\theta(\textbf{x}_{0:T}):=p(\textbf{x}_T)\prod_{t=1}^T(\textbf{x}_{t-1}|\textbf{x}_t).
\end{equation}
The training is done using variational lower bound (ELBO), shown as,
\begin{equation}
    \begin{aligned}
        \mathbb{E}[-\log p_\theta(\textbf{x}_0)]\leq\mathbb{E}_q\left[-\frac{p_\theta(\textbf{x}_{0:T})}{q(\textbf{x}_{1:T}|\textbf{x}_0)}\right] \\
        =\mathbb{E}_q\left[-\log p(\textbf{x}_T)-\sum_{t\geq 1}\log\frac{p_\theta(.)}{q(.)}\right] := \mathcal{L}
    \end{aligned}
\end{equation}
where $\mathcal{L}$ is the loss function. This loss function can be further simplified as 
\begin{equation}
\label{simple}
    \mathcal{L}_\mathcal{D}(\theta) := \mathbb{E}_{t,\textbf{x}_0,\epsilon}\left[||\epsilon-\epsilon_\theta(\sqrt{\alpha_t}\textbf{x}_0+\sqrt{1-\alpha_t}\epsilon,t)||^2\right]
\end{equation}
Now, when we add conditioning $c$ to this, the modified loss function can be written as
\begin{equation}
\label{controlnet}
        \mathcal{L}_{CN}(\theta) := \mathbb{E}_{\textbf{x}_0,t,c,\epsilon}\left[||\epsilon-\epsilon_\theta(\textbf{x}_t,t,c)||^2\right]
\end{equation}
\begin{figure}[t]
    \centering
\includegraphics[width=1\linewidth]{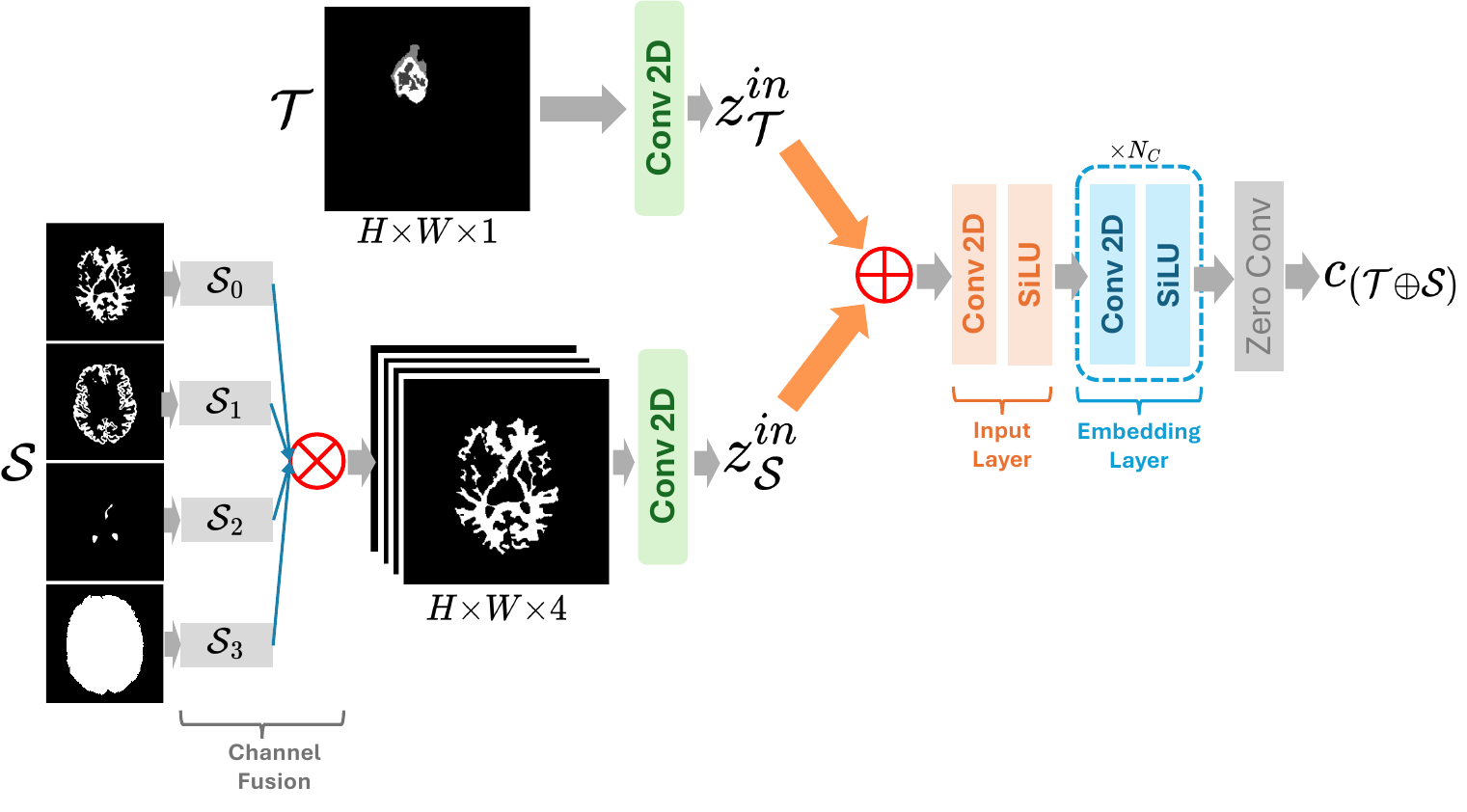}
\caption{\textbf{Tumor+Structure Aggregation (TSA) module.} The brain structure masks—WMT, CGM, LV, and BM—are fused with the tumor mask to create a unified representation, which serves as a conditional control for the diffusion model.}.
%\saum{the S0 S1 S2 S3 is too small to see. Maybe stack the S images vertically so that you have space to increase font size of S0-S3} \saum{increase font size of input layer and embedding layer texts}
\label{fig:tsa}
\end{figure}
\subsection{Tumor+Structure Aggregation} 
\label{sec:tsa}
%\pp{Structure-aware topology - dont think this is the right terminology} 
The purpose of the TSA module is to combine the brain anatomy with the tumor structure into a unified representation suitable for conditional control. The tumor masks are represented as $\mathcal{T}\in\mathbb{R}^{H \times W \times 1}$. The WMT, CGM, LV and BM are collectively represented as $\mathcal{S}\in\mathbb{R}^{H \times W \times 4}$. The channel dimension is $4$ as there are four anatomical masks, with each contributing $1$ channel. The fusion of $\mathcal{T}$ and $\mathcal{S}$ is denoted as $c_{(\mathcal{T}\otimes\mathcal{S})}$. This control $c_{(\mathcal{T}\otimes\mathcal{S})}$ is used for conditioning a diffusion model which is trained in a ControlNet-like \cite{zhang2023adding} manner.

\myparagraph{Brain anatomy.} In addition to the tumor, we use several other anatomically important structures namely White Matter Tracts (WMT), Cortical gray Matter (CGM), Lateral Ventricles (LV), and Brain Masks (BM). The detailed explanation about the generation of the structures are discussed in~\cref{implementation}. In~\cref{fig:structure}, we show examples of the different structures. More examples are shown in Supplementary (Figure \ref{fig:esupp2}).
%\saum{mention supp section number}. \saum{this paragraph seems like a repeat of the prev one.}

\myparagraph{Fusion.} In~\cref{fig:architecture}, we provide a brief overview of the TSA module. The different anatomies $\mathcal{T}$ and $\mathcal{S}$ are aggregated and fed as a control to the U-Net. The \textcolor{red}{$\oplus$} operator aggregates $\mathcal{S}$ across the channel dimension, and this is then fused with the $\mathcal{T}$ using the \textcolor{red}{$\otimes$} operator. Here, $\oplus$ and $\otimes$ denote element-wise sum and product respectively.
%\saum{do you wanna mention $\oplus$ and $\otimes$ denote element-wise sum and product respectively?} 
In~\cref{fig:tsa}, we show the inner workings of the TSA module. %the detailed structural components of the TSA module. 
A channel fusion module aggregates the different structures $\mathcal{S}_i\in\mathbb{R}^{H{\times}W{\times}1}$ into one unified feature representation $\mathcal{S}$. Both $\mathcal{T}$ and $\mathcal{S}$ are fed to a 2D convolution layer, $Conv2D$, to generate the input feature representations $z^{in}_\mathcal{T}$ and $z^{in}_\mathcal{S}$. These representations are then concatenated to a fused representation $z^{in}$. This is a weighted fusion where $z^{in} = \lambda_1 \cdot z_\mathcal{T}^{in}+\lambda_2 \cdot z_\mathcal{S}^{in}$ where $\lambda_1>>\lambda_2$. This is then fed to a input layer to generate $\hat{z}=SiLU(Conv2D(z^{in}))$. This is further encoded using the embedding layer to generate $\hat{z}_{i+1}=\sum^{N_C}_i SiLU(Conv2D(\hat{z_i}))$, where SiLU denotes the Sigmoid Linear Unit activation function. Then, a $ZeroConv$ layer, $\mathcal{C}_0,$ is applied to the output embedding to generate the final control embedding:

\begin{equation}
\label{control}
c_{(\mathcal{T}\otimes\mathcal{S})}:=\mathcal{C}_0(\hat{z}_{N_C})
\end{equation}

Here, $\hat{z}_{N_C}$ is the feature of the last layer of the embedding layer. 
%\saum{you should provide one sentence each about why you performed each operation. just to give intuition to the reader of why these steps are needed}
\begin{figure}[t]
    \centering
\includegraphics[width=1\linewidth]{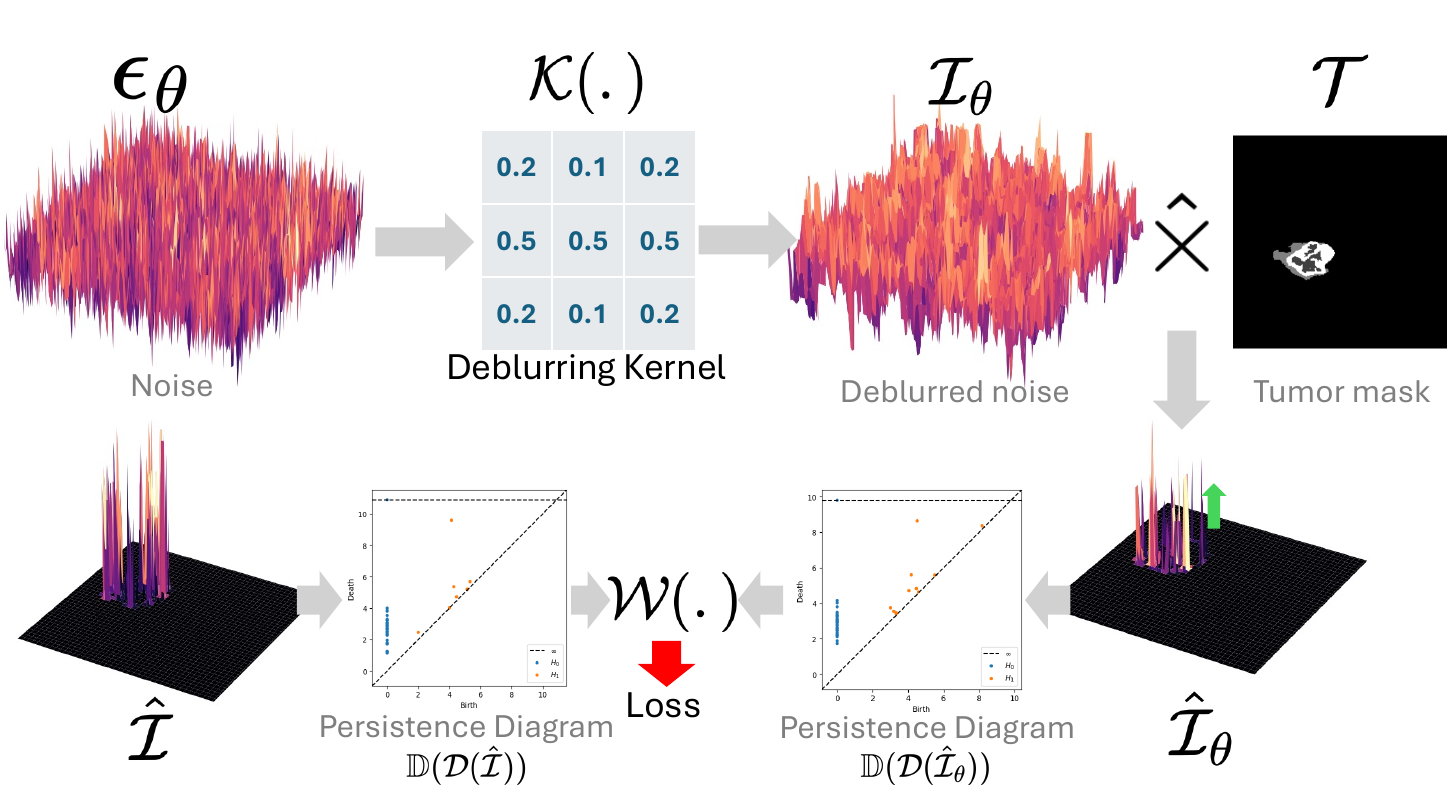}
\caption{\textbf{Tumor-Guided Anatomy Preserving (TGAP) module.} Predicted noise from the diffusion model is first deblurred, followed by masking of the tumor region. The PD is then computed from the mask, followed by loss calculation}. 
%\saum{use some english text in figure like Persistence Diagram D, etc (if you have time/space) \saum{loss arrow can be red like fig2}}
    \label{fig:tgap}
\end{figure}
\subsection{Topology-guided Anatomy Preservation}
\label{sec:tgap}
In this subsection, we describe the Topology-guided Anatomy Preservation (TGAP) module which aims to maintain the spatial consistency and heterogeneity
%\saum{is it more than just spatial correctness? Heterogeneity / irregular details, too?} 
of the tumor structure in the generated image. The condition $c_{(\mathcal{T}\otimes\mathcal{S})}$ from~\cref{control} is used to control the diffusion model. During training, the denoising model predicts the noise $\epsilon_\theta(\textbf{x}_t,t,c)$. Here, we can substitute $c$ with $c_{(\mathcal{T}\otimes\mathcal{S})}$ and the representation becomes $\epsilon_\theta(\textbf{x}_t,t,c_{(\mathcal{T}\otimes\mathcal{S})})$. 

To the predicted noise $\epsilon_\theta(\textbf{x}_t,t,c_{(\mathcal{T}\otimes\mathcal{S})})$, we first apply a deblurring kernel $\mathcal{K}(.)$ to obtain meaningful signal from the noise.
%\saumtext{continue this sentence....why is the deblurring kernel applied? To get some meaningful signal from the noise?}. 
We use $\mathcal{I}_\theta\in\mathbb{R}^{H\times W}$ to denote this deblurred predicton.
%From now on the deblurred noise is represented as $\mathcal{I}_\theta\in\mathbb{R}^{H\times W}$. 
%\saumtext{We further constrain $\mathcal{I}_\theta$ to focus only on the tumor region by multiplying \saum{is multiplying and cropping operation different?} it with the tumor mask.}
We further constrain $\mathcal{I}_\theta$ to focus only on the tumor region by multiplying it with the tumor mask.
%Now from this modified representation, the tumor masked region is cropped resulting 
This results in the cropped image $\hat{\mathcal{I}}_\theta:=\mathcal{I}_\theta\hat{\times}\mathcal{T}$, where $\hat{\mathcal{I}}_\theta\in\mathcal{R}^{H\times W}$ and $\hat{\times}$ is the cropping operator. The same operations are performed on the initial noise $\epsilon$ to obtain the representation $\hat{\mathcal{I}}$. Then, we compute persistence diagram $\mathbb{D}$ of both the images, $\mathcal{D}(\hat{\mathcal{I}}_\theta)$ and $\mathcal{D}(\hat{\mathcal{I}})$, represented as $\mathbb{D}(\mathcal{D}(\hat{\mathcal{I}}_\theta))$ and $\mathbb{D}(\mathcal{D}(\hat{\mathcal{I}}))$ respectively. Here, $\mathcal{D}(.)$ is the distance transform applied to the images. The distance transform is derived by applying soft-thresholding to the continuous pixel distribution in the tumor region, distinguishing it from the background.
%\saum{Question: Usually distance transform is applied on binary images, but you seem to apply it on continuous images I. I am not sure what distance transform is supposed to denote here?} 
Similar to previous topology loss calculations~\cite{hu2019topology}, we apply a Wasserstein distance~\cite{panaretos2019statistical} between the two diagrams. 

\textbf{Definition.} Given two diagrams $\mathbb{D}(P)$ and $\mathbb{D}(Q)$, the
$r$-th Wasserstein distance is defined as follows:
\begin{equation}
    \mathcal{W}\left(\mathbb{D}(P),\mathbb{D}(Q)\right)=\left(inf_{\gamma\in\tau}\sum_{x\in\mathbb{D}(P)}||x-\gamma(x)||^r\right)^\frac{1}{r}
\end{equation}
The motivation behind this loss is to guide the reverse process in a way that ensures each denoising step progressively recovers tumor details. Without this guidance, the MSE loss alone lacks the capacity to effectively restore fine details.
%\saum{need to add some sentences on the motivation behind this loss...guiding the reverse process in this manner ensures that each denoising step recovers tumor details in an iterative fashion ...something like that....and that without this guidance, the MSE loss alone is not powerful enough to recover details}
Here, $\tau$ represents all bijections from $\mathbb{D}(P)$ to $\mathbb{D}(Q)$.
\begin{table*}[t]
\small
\centering
\scalebox{0.8}{
\begin{tabular}{ccccccccccccc}
\hline
& \multicolumn{4}{c}{\textbf{PSNR($\uparrow$)}} & \multicolumn{4}{c}{\textbf{SSIM($\uparrow$)}} & \multicolumn{4}{c}{\textbf{MMD($\downarrow$)}}\\
\hline
& FLAIR & T1 & T1CE & T2 & FLAIR & T1 & T1CE & T2 & FLAIR & T1 & T1CE & T2 \\
\hline
\hline
% \multirow{11}{*}{\rotatebox{90}{BRATS-AG}} & 
SPADE\cite{park2019semantic} & \underline{63.09} & \underline{64.49} & \textbf{64.47} & \underline{63.78} & 0.1487 & 0.1852 & 0.2608 & 0.1131 & 2.0155 & 1.7417 & \textbf{2.0960} & 1.5210\\
% VQ-GAN \\
% VQ-VAE\cite{van2017neural} \\
DDPM\cite{ho2020denoising} & 57.91 & 59.09 & 59.60 & 58.15 & 0.1025 & 0.1196 & 0.1495 & 0.1006 & 8.9395 & 2.7057 & 6.2372 & 9.0479\\
LDM\cite{rombach2022high} & 57.62 & 55.32 & 59.83 & 60.05 & 0.0474 & 0.0530 & 0.0831 & 0.0775 & 6.7483 & 18.7172 & 4.3754 & 4.4354\\
% \cline{2-14}
\hline
Brain & 62.91 & 63.91 & 60.95 & 63.57 & \underline{0.3036} & \underline{0.3505} & \underline{0.2743} & \underline{0.2893} & 1.8011 & 1.2308 & 6.2626 & 0.9685\\
Tumor & 62.82 & 61.12 & 60.87 & 63.03 & 0.2733 & 0.2769 & 0.2428 & 0.2807 & 1.0283 & \underline{1.1183} & 5.0314 & 0.6332\\
WMT & 62.07 & 60.84 & 61.29 & 62.32 & 0.1694 & 0.2257 & 0.1753 & 0.1566 & \underline{0.4842} & 1.5697 & 3.5701 & \underline{0.5285}\\
CGM & 62.13 & 60.66 & 61.40 & 62.19 & 0.1670 & 0.2197 & 0.1680 & 0.1539 & 0.4881 & 1.7301 & 3.5384 & 0.5498\\
LV & 62.35 & 60.85 & 61.77 & 62.38 & 0.1763 & 0.2123 & 0.1781 & 0.1459 & \textbf{0.4135} & 1.7137 & \underline{3.1101} & 0.5924\\
% \cline{2-14}
\hline
% 65.40 & 65.36 & 62.06 & 66.73 & 0.3554 & 0.4396 & 0.3228 & 0.3860 & 1.3251 & 1.0008 & 4.8089 & 0.3976\\
Ours & \textbf{65.40} & \textbf{65.36} & \underline{62.06} & \textbf{66.73} & \textbf{0.3554} & \textbf{0.4396} & \textbf{0.3228} & \textbf{0.3860} & 1.3251 & \textbf{1.0008} & 4.8089 & \textbf{0.3976}\\
% & \textbf{65.40} & \textbf{65.36} & \textbf{62.07} & \textbf{66.73} & \textbf{0.3554} & \textbf{0.4396} & \textbf{0.3228} & \textbf{0.3860} & 1.3251 & \textbf{1.0008} & 4.8089 & \textbf{0.3976}\\
\hline
% \multirow{11}{*}{\rotatebox{90}{BRATS-Met}} & SPADE \\
% &VQ-GAN \\
% &VQ-VAE \\
% &DDPM \\%& 57.91 & 59.09 & 59.60 & 58.15 & 0.1025 & 0.1196 & 0.1495 & 0.1006 & 8.9395 & 2.7057 & 6.2372 & 9.0479\\
% &LDM \\%& 57.62 & 55.32 & 59.83 & 60.05 & 0.0474 & 0.0530 & 0.0831 & 0.0775 & 6.7483 & 18.7172 & 4.3754 & 4.4354\\
% \cline{2-14}
% &Brain \\%& 62.91 & 63.91 & 60.95 & 63.57 & 0.3036 & 0.3505 & 0.2743 & 0.2893 & 1.8011 & 1.2308 & 6.2626 & 0.9685\\
% &Tumor \\%& 62.82 & 61.12 & 60.87 & 63.03 & 0.2733 & 0.2769 & 0.2428 & 0.2807 & 1.0283 & 1.1183 & 5.0314 & 0.6332\\
% % Brain+Tumor & 64.57 & 64.02 & 61.58 & 64.09 & 0.2973 & 0.3817 & 0.2813 & 0.2710 & 1.6659 & 1.4238 & 5.6044 & 1.0100\\
% &WMT & \\%63.26 & 64.04 & 60.88 & 65.10 & 0.3095 & 0.4005 & 0.3150 & 0.3323 & 1.4727 & 1.1755 & 6.2787 & 0.6446\\
% &CGM \\%& 62.14 & 61.55 & 60.57 & 63.22 & 0.2361 & 0.2969 & 0.2882 & 0.2938 & 0.8969 & 2.0878 & 6.2469 & 0.6359\\
% &LV \\%& 60.95 & 59.51 & 61.00 & 62.38 & 0.2329 & 0.1977 & 0.2330 & 0.2030 & 2.3818 & 3.6773 & 3.7216 & 1.0556\\
% \cline{2-14}
% &Ours \\
% % Ours & \textbf{63.32} & 63.91 & 60.56 & \textbf{66.16} & \textbf{0.3639} & 0.3977 & 0.3133 & \textbf{0.4319} & 3.0885 & 2.1963 &7.6704 & \textbf{0.1486}\\
% \hline
\end{tabular}
}
\caption{\textbf{Image Quality Assessment on BraTS-AG dataset}: We report PSNR, SSIM and MMD metrics for different GAN and diffusion based baselines and compare with our BrainMRDiff method. Best results in \textbf{bold} and second best in \underline{underline}.\label{tab1} %\saum{ICCV template is to have table captions below the table}
}
\end{table*}

\begin{table*}
\small
\centering
\scalebox{0.8}{
\begin{tabular}{ccccccccccc}
\hline
 & \multicolumn{3}{c}{\textbf{PSNR($\uparrow$)}} & \multicolumn{3}{c}{\textbf{SSIM($\uparrow$)}} & \multicolumn{3}{c}{\textbf{MMD($\downarrow$)}}\\
\hline
& T1C & T1N & T2F & T1C & T1N & T2F & T1C & T1N & T2F\\
\hline
\hline
% \multirow{11}{*}{\rotatebox{90}{BRATS-Met}} & 
SPADE & \underline{61.94} & \underline{63.18} & \textbf{64.59} & 0.1568 & 0.1512 & \underline{0.1639} & \underline{4.1250} & \underline{1.4909} & \underline{1.1038}\\
% VQ-GAN \\
% VQ-VAE \\

DDPM & 58.47 & 60.13 & 58.93 & 0.1056 & 0.1245 & 0.1019 & 8.1038 & 1.5968 & 5.0377\\
LDM & 56.40 &
54.19 &
55.86 &
0.0682 &
0.0353 &
0.0504 &
11.5967 &
17.4827 &
12.1118 \\
\hline
Brain & 58.59 & 59.96 & 59.02 & 0.1590 & 0.1585 & 0.1248 & 7.3282 & 2.1151 & 4.2226\\
Tumor & 58.84 & 59.87 & 58.94 & 0.1648 & 0.1553 & 0.1207 & 6.9952 & 2.3973 & 4.3585\\
WMT & 58.61 & 59.97 & 59.09 & 0.1596 & 0.1581 & 0.1248 & 7.1652 & 2.1548 & 4.1261\\
CGM & 58.78 & 59.95 & 59.05 & 0.1628 & \underline{0.1588} & 0.1236 & 6.8161 & 2.3123 & 4.2388\\
LV & 58.92 & 59.94 & 59.00 & \underline{0.1665} & 0.1579 & 0.1235 & 6.7353 & 2.2444 & 4.3641\\
\hline
% 62.14 & 63.80 & 64.35 & 0.2040 & 0.2072 & 0.1837 & 2.9376 & 0.2307 & 1.0702\\
Ours & \textbf{62.14} & \textbf{63.80} & \underline{64.35} & \textbf{0.2040} & \textbf{0.2072} & \textbf{0.1837} & \textbf{2.9376} & \textbf{0.2307} & \textbf{1.0702}\\
\hline
\end{tabular}
}
\caption{\textbf{Image Quality Assessment on BraTS-Met dataset}}\label{tab2}
\end{table*}

\subsection{End-to-end Training}
\label{sec:training}
We first train a diffusion model with different sequences using~\cref{simple}. We then freeze the parameters of this trained diffusion model and condition it with the control from the TSA module as represented in~\cref{control}. Further, we use the loss function from~\cref{controlnet}. This is the TSA loss function $\mathcal{L}_\text{TSA}$ and can be represented as 
\begin{equation}
\label{label_tsa}
    \mathcal{L}_\text{TSA} := \mathbb{E}_{\textbf{x}_0,t,c_{(\mathcal{T}\otimes\mathcal{S})},\epsilon}\left[||\epsilon-\epsilon_\theta(\textbf{x}_t,t,c_{(\mathcal{T}\otimes\mathcal{S})})||^2\right]
\end{equation}
We then calculate the Mean Squared Error (MSE) loss between the initial noise $\epsilon$ and the predicted noise $\epsilon_\theta$, shown as $\mathcal{L}_\text{MSE}=\frac{1}{n}\sum_i^n\left(\mathcal{I}_\theta-\mathcal{I}\right)^2$. And the loss function of the TGAP module, $\mathcal{L}_\text{TGAP}$ can be represented as 
\begin{equation}
\label{label_tgap}
    \mathcal{L}_\text{TGAP} = \mathcal{W}\left(\mathbb{D}(P), \mathbb{D}(Q)\right)
\end{equation}
Hence, the combined final loss function, $\mathcal{L}_\text{final}$, is a combination of the MSE and TGAP loss functions. From~\cref{label_tsa} and~\cref{label_tgap}, $\mathcal{L}_\text{final}$ can be represented as $\mathcal{L}_\text{final}=\mathcal{L}_\text{MSE}+\lambda\mathcal{L}_\text{TGAP}$. Here, $\lambda$ is the weighting parameter.
%\subsection{Inference}

% \begin{table*}
%   \centering
%   \begin{tabular}{@{}lc@{}}
%     \toprule
%     Method & Frobnability \\
%     \midrule
%     Theirs & Frumpy \\
%     Yours & Frobbly \\
%     Ours & Makes one's heart Frob\\
%     \bottomrule
%   \end{tabular}
%   \caption{Results.   Ours is better.}
%   \label{tab:example}
% \end{table*}

\begin{table}
\small
\centering
\begin{tabular}{lll}
\hline
DSC ($\uparrow$) & Brain & Tumor\\
\hline
\hline
DDPM & \underline{0.6411} & 0.0727\\
ControlNet & \textbf{0.6412} & \underline{0.5689}\\
\hline
Ours & \textbf{0.6412} & \textbf{0.5736}\\
\hline
\end{tabular}
\caption{\textbf{Segmentation results.} We report DSC score for BM and tumor segmentation for BrainMRDiff and compare with baselines such as DDPM and ControlNet. Best results in \textbf{bold} and second best in \underline{underline}.}\label{tab_segmentation}
\end{table}

%% file: sections/4_results.tex
\section{Experiments}
\label{results}
\subsection{Datasets and Implementations}
\label{datasets}
\label{implementation}
%For experimentation, we use the 
We validate our method on the Brain Tumor Segmentation-2021 (BraTS-2021)~\cite{baid2021rsna} and the Brain Tumor Segmentation - Metastasis (BraTS-Met)\cite{moawad2024brain} datasets. The BraTS-AG dataset comprises of patients with adult glioma (glioblastoma and astrocytoma) segmentation on pre-treatment MRI. We partition the dataset into training (1022), validation (113), and testing (116) subsets, ensuring that MGMT and OS status are available for all cases in the test set. For this dataset, we utilize all four MRI sequences—FLAIR, T1, T1CE, and T2—for experimentation. The BraTS-Met dataset focuses on brain metastasis segmentation on pre-treatment MRI. For this dataset, we conduct experiments using T1C, T1N, and T2F sequences. The training, validation and testing set contains 398, 98 and 142 patients respectively.
%Similarly, this dataset is also divided into training, validation and testing sets. 
%For MGMT prediction, we use a MLP classifier layer. For survival status we use ...
%\subsection{Implementation}

For both datasets, we standardize the image spacing to $[1.5, 1.5, 1.0]$ and resize the slices to $(128\times128)$ pixels. The architectures are implemented using PyTorch~\cite{paszke2019pytorch} and MONAI Generative~\cite{cardoso2022monai,pinaya2023generative}. Anatomical structures of the brain are generated using SynthSeg~\cite{billot2023synthseg}, while tumor masks are created by merging the nested tumor regions (WT, TC, ET) into a  unified mask. Tumor segmentation for the generated images is performed using a trained nnU-Net. Model training is conducted with the Adam optimizer, employing a lr of $2.5e-5$ and a batch size of 2 (all experiments on a Quadro RTX 8000 GPU with 48 GB of memory).
%\subsection{Evaluation Metrics} For the image quality assessment (IQA), we use Peak Signal-to-Noise Ratio (PSNR)~\cite{}, Structural Similarity Index Measure (SSIM)~\cite{} and Maximum Mean Discrepancy (MMD)~\cite{}. For segmentation, we use Dice Similarity Coefficient (DSC), Average Surface Distance (ASD) and Jaccard Coefficient (JC). For MGMT classification, we report Accuracy, Area-Under-Curve (AUC) score, F1-score, Precision and Recall. For survival analysis, we report Concordance-index (c-index).
\subsection{Quantitative Analysis}
In this subsection, we present the quantitative evaluation of our proposed methods. Our experimental framework encompasses two primary tasks: Image Quality Assessment (IQA) and Segmentation. Additionally, we provide ablation studies to analyze the contributions of various components within our methodology. In our experimental setup, anatomy masks obtained from SynthSeg are used as input to generate MR sequences. For IQA experiments, we compute image quality metrics for both generated and real images. In segmentation experiments, BM and tumor masks are extracted and compared with their real counterparts. The experimental setup remains consistent across both datasets. DDPM and LDM use noise as input, while SPADE and ControlNet (Brain) use BM. Other ControlNet models take tumor, WMT, CGM, or LV as inputs to generate MR sequences.\\
\textbf{Image Quality Assessment.}
%For both experimental settings, 
%\mb{Here, ...}
For both the datasets, we benchmark our method against standard generative models, including the GAN-based SPADE~\cite{park2019semantic}, diffusion-based DDPM~\cite{ho2020denoising}, and LDM~\cite{rombach2022high}. Further, we compare our approach with ControlNet~\cite{zhang2023adding}, which is trained with different anatomical structures as control mechanisms. \cref{tab1} presents the IQA results for the BraTS-AG dataset. Our method, BrainMRDiff, demonstrates superior performance over all baselines across all sequences in terms of SSIM scores. Additionally, it outperforms baselines in three out of four sequences for the PSNR score and in two out of four sequences for the MMD score. Specifically, BrainMRDiff achieves a combined PSNR score of 64.89±1.72, surpassing SPADE, which records a combined PSNR score of 63.96±0.58. The combined score is the mean±std. across all the different sequences. Moreover, BrainMRDiff attains a combined SSIM score of 0.37±0.05, with the second-best baseline being ControlNet trained with BM as control, yielding a combined SSIM score of 0.30±0.03. However, for the MMD score, BrainMRDiff attains a combined value of 1.88±1.72, whereas ControlNet trained with LV achieves a superior score of 1.46±1.08. Notably, ControlNets trained on individual anatomical structures (WMT, CGM, and LV) outperform BrainMRDiff in terms of the MMD metric. \cref{tab2} reports the IQA results for the BraTS-Met dataset. BrainMRDiff outperforms all baseline models across all sequences for SSIM and MMD scores and outperforms in two out of three sequences for the PSNR score. Furthermore, it achieves the highest combined performance across all metrics. Specifically, BrainMRDiff attains a combined PSNR score of 63.43±0.94, exceeding SPADE, which records 63.24±1.08. For the SSIM metric, BrainMRDiff achieves a combined score of 0.20±0.01, with the second-best baseline being ControlNet trained with LV, which obtains 0.15±0.02. Regarding the MMD metric, BrainMRDiff achieves a score of 1.41±1.13, outperforming SPADE, which records a score of 2.24±1.34. \\
These findings demonstrate the efficacy of BrainMRDiff in enhancing image quality across multiple evaluation metrics across datasets.

\begin{table}[t]
\small
\centering
\begin{tabular}{cccccc}
\hline
- &  \multicolumn{4}{c}{\textbf{SSIM}($\uparrow$)} & \textbf{DSC}($\uparrow$)\\
\hline
- & FLAIR & T1 & T1CE & T2 & Tumor\\
\hline
\hline
TSA & \underline{0.3377} & \underline{0.4281} & 0.2552 & \underline{0.3123} & \underline{0.5668}\\
TGAP & 0.3375 & 0.4212 & \underline{0.2777} & 0.2855 & 0.2462\\
\hline
Ours & \textbf{0.3554} & \textbf{0.4396} & \textbf{0.3228} & \textbf{0.3860} & \textbf{0.5736}\\
\hline
\end{tabular}
\caption{\textbf{Ablation results.} We show the results for different component of BrainMRDiff namely TSA and TGAP for IQA and BM segmentation tasks. We report SSIM score for IQA and DSC score for segmentation. Best results in \textbf{bold} and second best in \underline{underline}.}\label{tab_ablation}
\end{table}

\begin{figure}
    \centering
\includegraphics[width=1\linewidth]{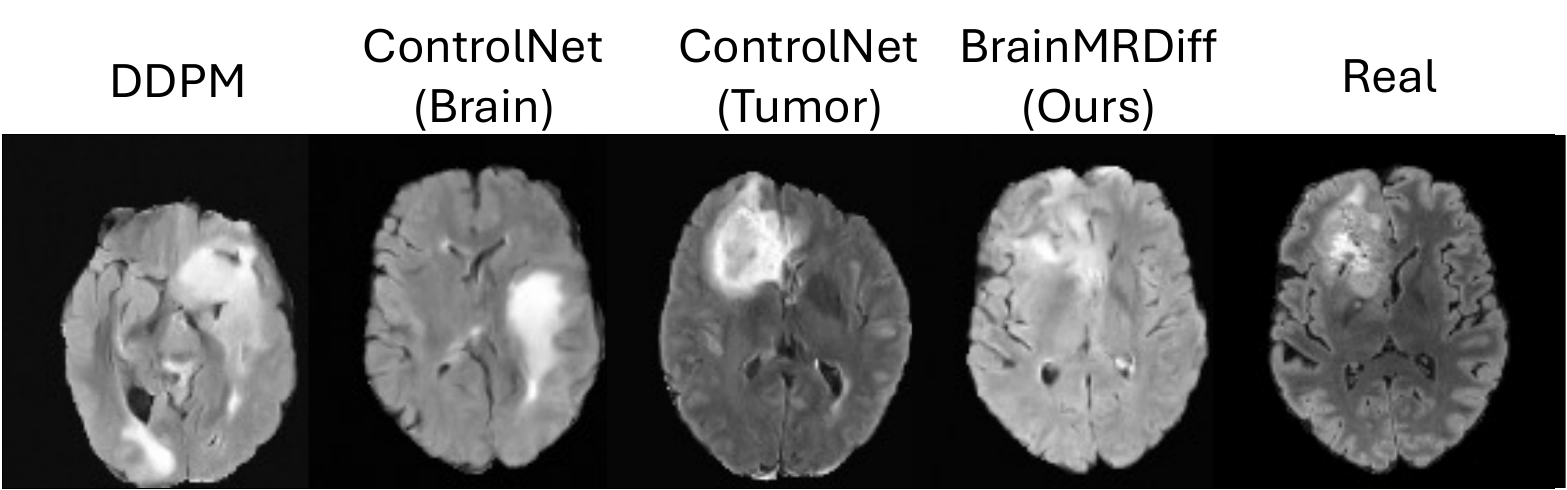}
    \caption{\textbf{Comparison with baselines.} A generated image from BrainMRDiff is compared with those from methods like DDPM and ControlNet. BrainMRDiff achieves superior anatomical consistency, maintaining both brain structures and tumor topology with high fidelity, closely resembling the real MRI scan. }
    %\pp{Moinak- can you get gagan's interpretation on these to demonstrate why yours is best?}} \gs{updated the caption}
    \label{fig:qualitative1}
\end{figure}

\myparagraph{Segmentations.} For segmentation, we compare our results against DDPM and ControlNet for both brain mask (BM) and tumor segmentation tasks. In this evaluation, ControlNet is trained separately on BM and tumor segmentations. The BMs for the generated images are obtained by thresholding pixel values above a predefined threshold, while the tumor segmentations for the generated images are derived from a pre-trained nnU-Net (implementation details provided in~\cref{implementation}). The segmentation results are summarized in~\cref{tab_segmentation}, where we report the combined Dice Similarity Coefficient (DSC) score across all sequences. Our findings indicate that BrainMRDiff outperforms DDPM (0.6411) in BM segmentation while achieving comparable performance to ControlNet (0.6412). This result suggests that for a single, large anatomical structure (in terms of pixel or voxel count), ControlNet is sufficient for generating accurate segmentation masks. However, when evaluating tumor segmentation, BrainMRDiff demonstrates superior performance over both baselines. Specifically, BrainMRDiff achieves a DSC score of 0.5736, surpassing ControlNet trained on tumor masks (0.5689) and significantly outperforming DDPM, which exhibits a considerably lower DSC score of 0.0727.\\
These results highlight the robustness of BrainMRDiff in handling complex segmentation tasks, particularly in scenarios involving smaller and more heterogeneous anatomical structures, such as tumors.\\
\begin{figure}[t]
    \centering
\includegraphics[width=.8\linewidth]{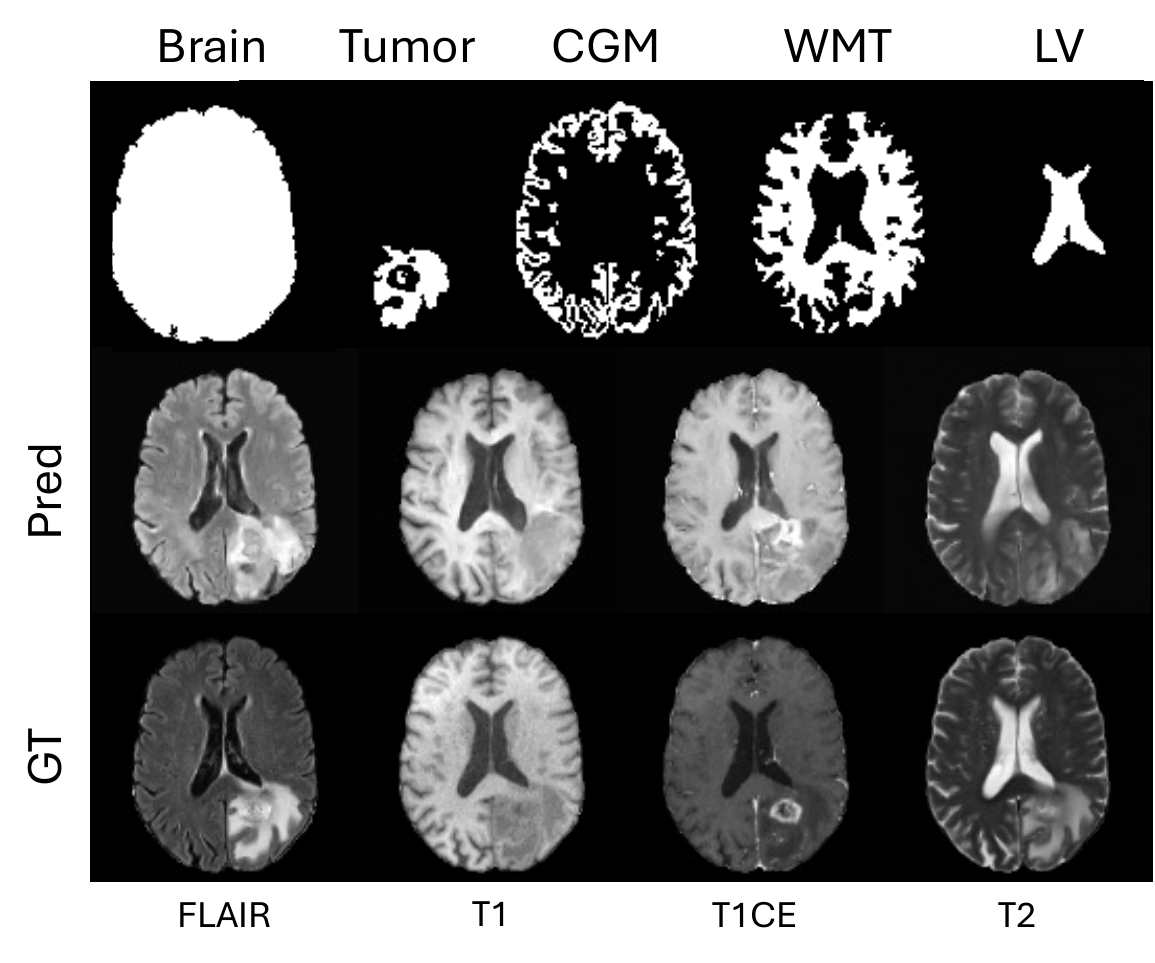}
    \caption{
    %\textbf{Additional Results.} We show an example of the generated sequence of our method along with the real images and the different tumor and anatomy segmentation masks.}
    Row 1: Tumor and Structure masks, Row 2: Generated multi-parametric MRI, Row 3: Ground Truth MRI.}
    \label{fig:qualitative2}
\end{figure}
\begin{table*}[t]
\small
\centering
\scalebox{0.8}{
\begin{tabular}{cccccc}
\hline
- & Bal. Ac. ($\uparrow$) & F1 ($\uparrow$) & Precision ($\uparrow$) & Recall ($\uparrow$) & c-index ($\uparrow$)\\
\hline
\hline
Original & \textit{65.75±3.35} & \textit{65.26±3.91} & \textit{67.15±3.12} & \textit{65.68±3.45} & \textit{0.65±0.02}\\
DDPM & \underline{53.23±3.83} & \underline{53.47±4.42} & \underline{54.00±4.03} & \underline{53.67±4.47} & 0.54±0.05\\
ControlNet & 49.82±11.89 & 47.64±11.19 & 51.21±13.18 & 48.58±11.20 & \underline{0.57±0.03}\\
\hline
Ours & \textbf{65.08±5.88} & \textbf{65.20±5.51} & \textbf{66.73±6.72} & \textbf{65.68±5.65} & \textbf{0.67±0.03}\\
\hline
\end{tabular}
}
\caption{\textbf{Clinical applications.} We show results for MGMT classification and survival analysis. We report Balanced Accuracy, F1-score, Precision and Recall for MGMT classification and c-index for survival analysis. Best results in \textbf{bold}, second best in \underline{underline} and the results for the original images in italics.}\label{tab_ca}
\end{table*}
\myparagraph{Ablations.} For the ablation experiments, we present the results of the two primary components of our proposed method, TSA and TGAP, in~\cref{tab_ablation}. We evaluate their performance on two distinct tasks: IQA and tumor segmentation. Our findings indicate that the combined approach, integrating both TSA and TGAP, delivers superior performance compared to the individual components in both tasks.
%in the IQA task. 
%Similarly, for tumor segmentation, the combined approach outperforms both the TSA and TGAP modules. 
%Specifically, the combined method achieves an SSIM score of 0.3760±0.0430, while the TSA module alone attains 0.3333±0.0623, and the TGAP module alone achieves 0.3305±0.0572. For the tumor segmentation task, BrainMRDiff attains a DSC of 0.5736, whereas the TSA module achieves 0.5668, and the TGAP module achieves 0.2462. 
Notably, the TSA module outperforms the TGAP module in both tasks. However, the combination of TSA and TGAP further enhances performance.
% For IQA, we report the Structural Similarity Index Measure (SSIM) scores, while for BM segmentation, we provide the Dice Similarity Coefficient (DSC) scores. 
%However, for the BM segmentation task, the combined approach performs comparably to the TGAP module alone.
%For the BM segmentation task, BrainMRDiff with the TGAP module attains a DSC of 0.6412 for FLAIR and T1 sequences and 0.6411 for T1CE and T2 sequences. In contrast, the TSA module achieves a DSC score of 0.6411 across all four sequences. 
%, while the TGAP module demonstrates superior performance in the BM segmentation task. This suggests that the TSA module is particularly effective for IQA, whereas the TGAP module contributes more substantially to BM segmentation. These findings show the complementary nature of the TSA and TGAP modules, highlighting their task-specific advantages.
\subsection{Qualitative Analysis}
For qualitative analysis, we compared the images generated by BrainMRDiff with those produced by different baseline models. In~\cref{fig:qualitative1}, we show a comparison of generated images from DDPM, ControlNet trained with BM, ControlNet trained with tumor segmentations, and BrainMRDiff, alondside the corresponding real FLAIR sequence from the BraTS-AG dataset. \underline{Neuroradiologist's (8 years exp) interpretation:} \textit{We observe that since DDPM lacks anatomical controls, it generates visually plausible MRI sequences but fails to preserve anatomical structures. In contrast, ControlNet introduces anatomical awareness, but its performance remains suboptimal in accurately capturing both brain and tumor morphology. BrainMRDiff, however, achieves superior fidelity by preserving both brain anatomical details and tumor topology, resulting in highly realistic and anatomically coherent MRI sequences.} In~\cref{fig:qualitative2}, we further illustrate real and BrainMRDiff-generated images along with the corresponding anatomical masks for two cases from the BraTS-AG dataset, showcasing our methods ability to generate high-quality and anatomically-accurate images. More examples are provided in the Supp. (Figs \ref{fig:supp1} and \ref{fig:esupp3}).
\begin{figure}
    \centering
\includegraphics[width=0.65\linewidth]{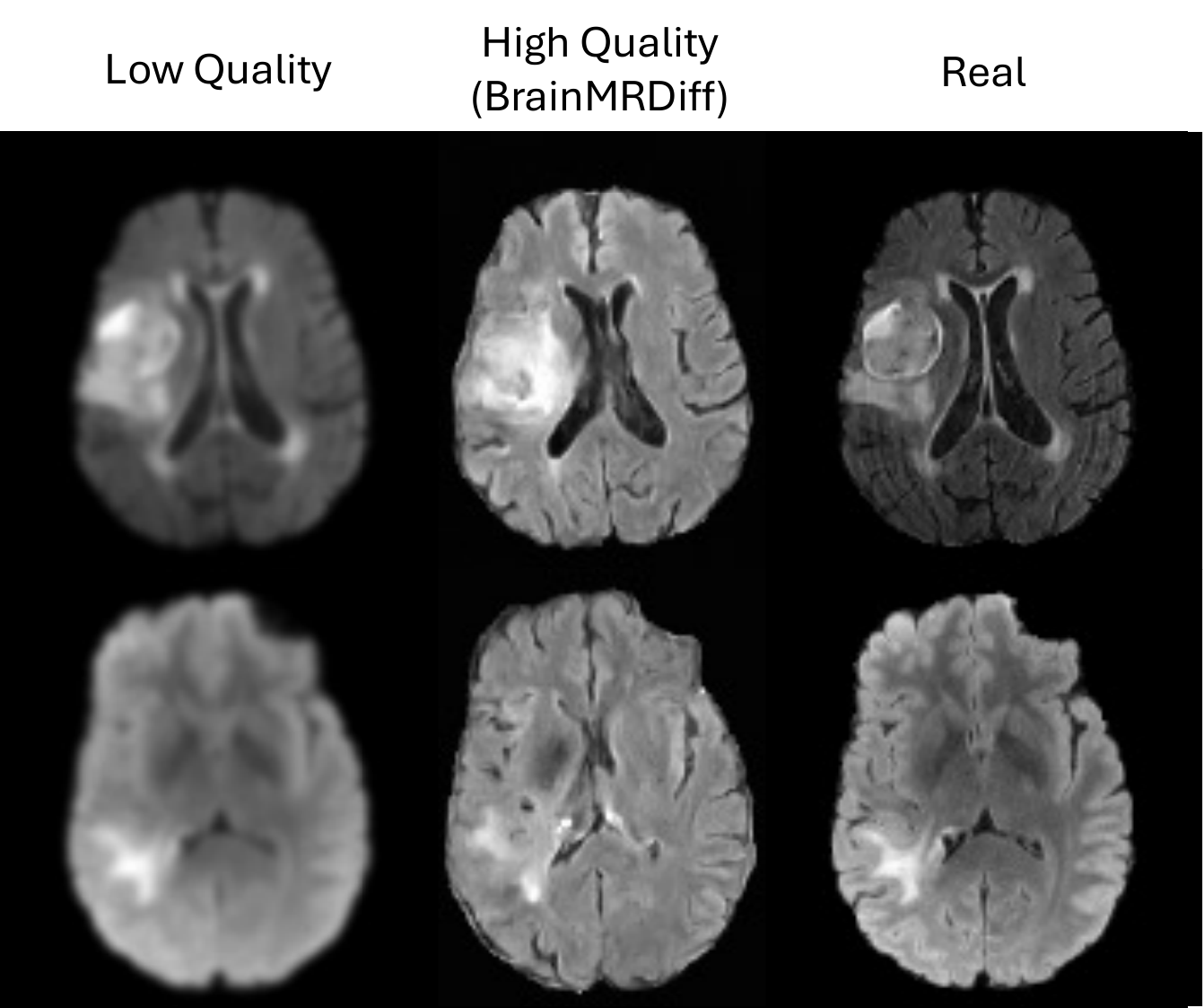}
    \caption{\textbf{High quality image generation.} We show the low quality image and the generated high quality scan from our method.}
    \label{fig:ca}
\end{figure}
\subsection{Clinical applications}
Beyond the IQA and segmentation experiments, we also evaluate our method on three clinically relevant tasks: High-quality image generation, MGMT status prediction and survival analysis.\\
\textbf{High quality image generation.} We artificially add gaussian noise to the images to create low quality images. From these low quality images, we obtain the anatomical structures and use them as control to BrainMRDiff. In~\cref{fig:ca}, we observe that BrainMRDiff generates a high-quality FLAIR sequence that closely resembles the real FLAIR sequence. \cref{tab_cax} presents the PSNR scores across varying noise levels. Notably, the generated image quality remains consistent regardless of noise intensity.\\
% and outperforms the low-quality images.}
%\pp{What about quantitative results?}\\
\textbf{MGMT status prediction.} We utilize tumor masks obtained from a pre-trained nnU-Net and extract radiomic features from 
%both 
the tumor regions in the generated images. 
%and the corresponding tumor segmentation masks. 
A multi-layer perceptron (MLP) classifier is trained on the radiomic features for MGMT status prediction in a 5-fold cross-validation setting. As shown in~\cref{tab_ca}, features derived from images generated by BrainMRDiff yield the highest classification performance compared to DDPM and ControlNet. Specifically, our method achieves an improvement of 11.85\% in balanced accuracy, and 11.73\% in F1-score. 
% , 12.73\% in precision, and 12.01\% in recall. 
Notably, BrainMRDiff achieves classification performance comparable to that of the original images, highlighting its potential for clinical application.\\
\textbf{Survival Analysis.} We further employ the same features for survival prediction, training a Cox Proportional Hazards model in a 5-fold cross-validation setting. BrainMRDiff achieves a concordance index of 0.67±0.03, marking an improvement of 0.1 over the second-best baseline, ControlNet. Interestingly, in this case, our approach even surpasses the prognostic performance of the original images.
% , highlighting its potential utility in prognostic modeling.
\begin{table}
\centering
\scalebox{0.8}{
\begin{tabular}{cccccc}
\hline
\multicolumn{2}{c}{$\sigma=0.5$} & \multicolumn{2}{c}{$\sigma=1$} & \multicolumn{2}{c}{$\sigma=2$}\\
\hline
LQ & Ours & LQ & Ours & LQ & Ours \\ 
\hline
\hline
65.41 & 66.16 & 65.23 & 67.33 & 64.99 & 66.73 \\
\hline
\end{tabular}
}
\caption{\textbf{High quality image generation (Quantitative).} We add noise to the real images (n=10) weighted by parameter $\sigma$. We report PSNR score for the Low Quality (LQ) images and our generated High Quality (HQ) images.}\label{tab_cax}
\end{table}

%0.5 
%0.4070(gt)
%65.78
%-
%0.4222(flair)
%66.16
%1
%0.4081(gt)
%65.42
%-
%0.4485(flair)
%67.33
%2
%0.3126(gt)
%64.91
%-
%0.3857(flair)
%66.73

%\subsection{Discussion}
%...
%In the future, we aim to extend this method to generated 3D MRI sequences. In addition to that, we aim to extend this method to generate high quality image generation for motion artifacts and generate MRI sequence from other modalities like CT, PET, etc. ...

%% file: sections/5_conclusion.tex
\section{Conclusion}
In conclusion, we introduce BrainMRDiff, an anatomy-guided diffusion model designed for the realistic generation of brain MRI sequences. 
% The proposed framework comprises two key modules: Tumor+Structure Aggregation (TSA) and Topology-Guided Anatomy Preservation (TGAP). The TSA module integrates anatomical structures, including the brain mask, white matter tracts (WMT), cortical gray matter (CGM), and lateral ventricles (LV), with tumor masks to provide anatomical control during the diffusion process. The TGAP module ensures the topological integrity of the tumor structure by computing the distance between the persistent diagrams of ground-truth and predicted segmentations. 
Experimental evaluations demonstrate the effectiveness of BrainMRDiff across multiple tasks, including image quality assessment, tumor segmentation, and clinically relevant applications such as MGMT promoter methylation status prediction and survival analysis. These results demonstrate the promising potential of our method for advancing real-world clinical applications in neuro-oncology, paving the way for more accurate and reliable brain MRI synthesis in clinical practice.\\
\textbf{Acknowledgments}  This research was supported by the National Institutes of Health (NIH) grants R21CA258493-01A1 and R01CA297843. The content is solely the responsibility of the authors and does not necessarily represent the official views of the National Institutes of Health.

%% file: supplementary.tex
%\section{Implementation}
\clearpage
\maketitlesupplementary

\begin{figure*}[hbt!]
    \centering
    \includegraphics[width=\linewidth]{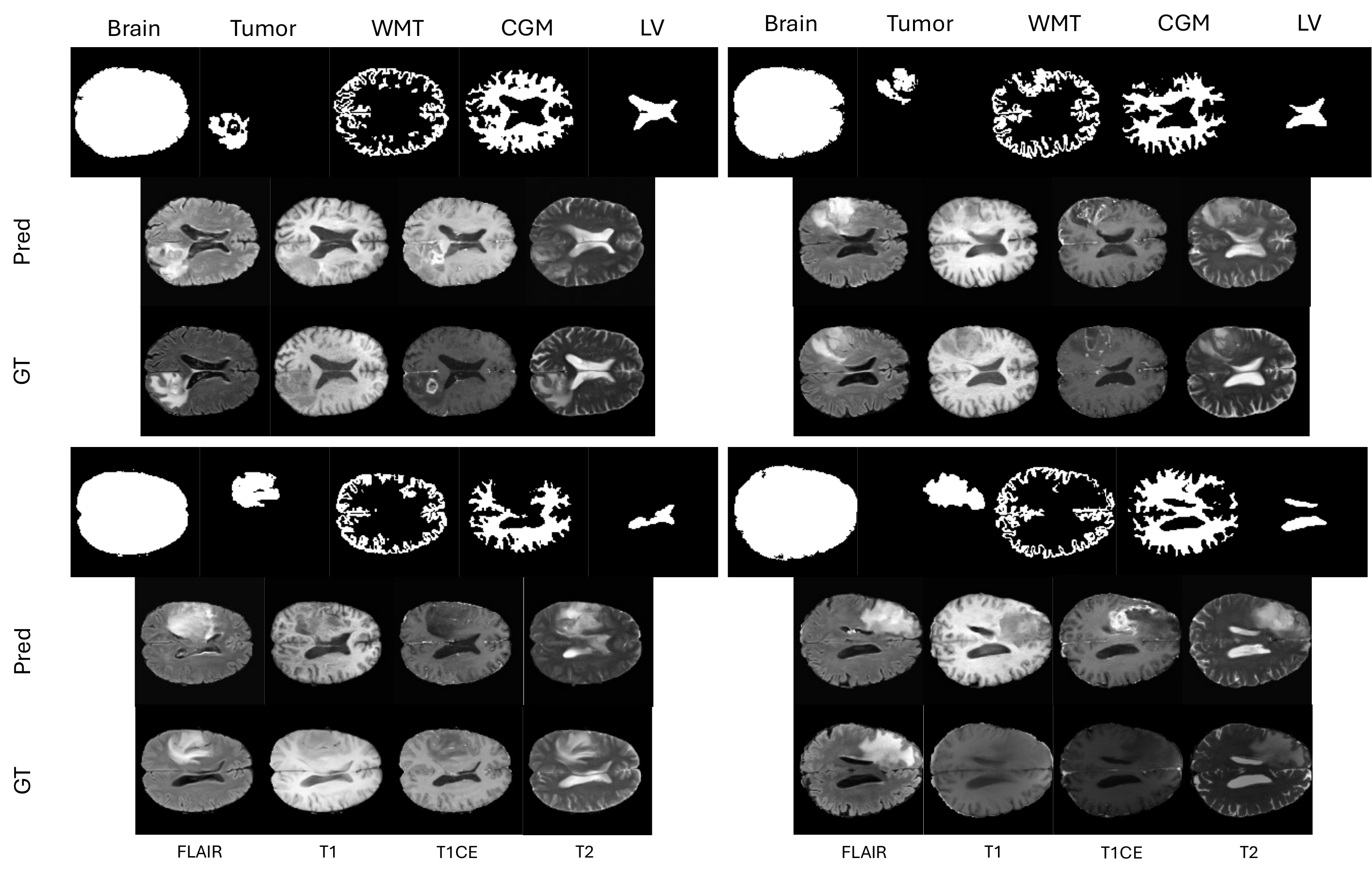}
    \caption{\textbf{Additional qualitative examples.} Example results generated by our method alongside ground truth images and segmentation masks. Row 1,4: Tumor and structure masks. Row 2,5: Generated multi-parametric MRI. Row 3,6: Ground truth MRI.}
    \label{fig:supp1}
\end{figure*}

\begin{figure*}
    \centering
    \includegraphics[width=1\linewidth]{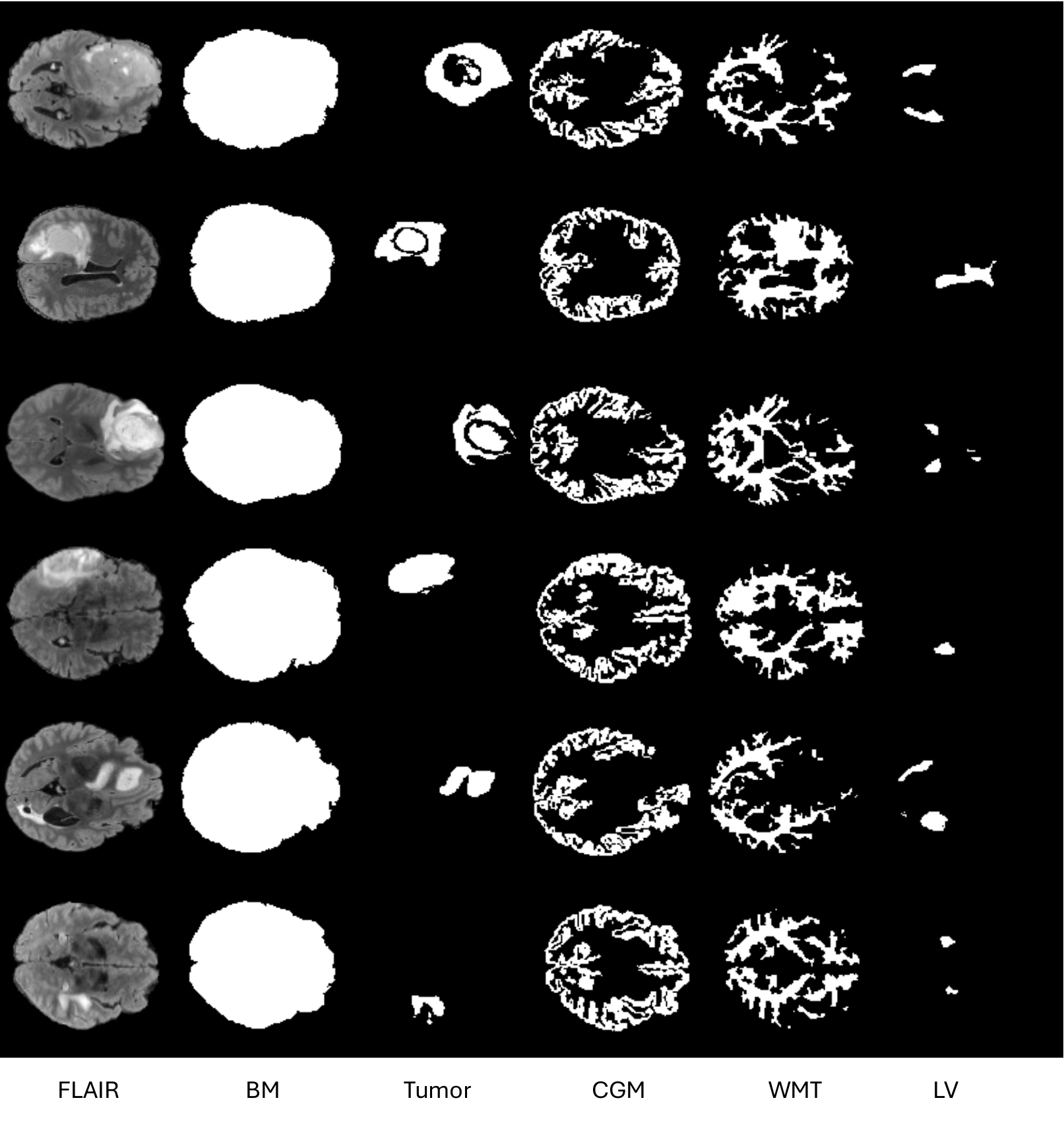}
    \caption{\textbf{Additional Examples of Anatomical Structure and Tumor Segmentations.} Six cases from the BraTS-AG dataset illustrating different anatomical structures and corresponding tumor segmentation masks.}
    \label{fig:esupp2}
\end{figure*}

\begin{figure*}
    \centering
    \includegraphics[width=1\linewidth]{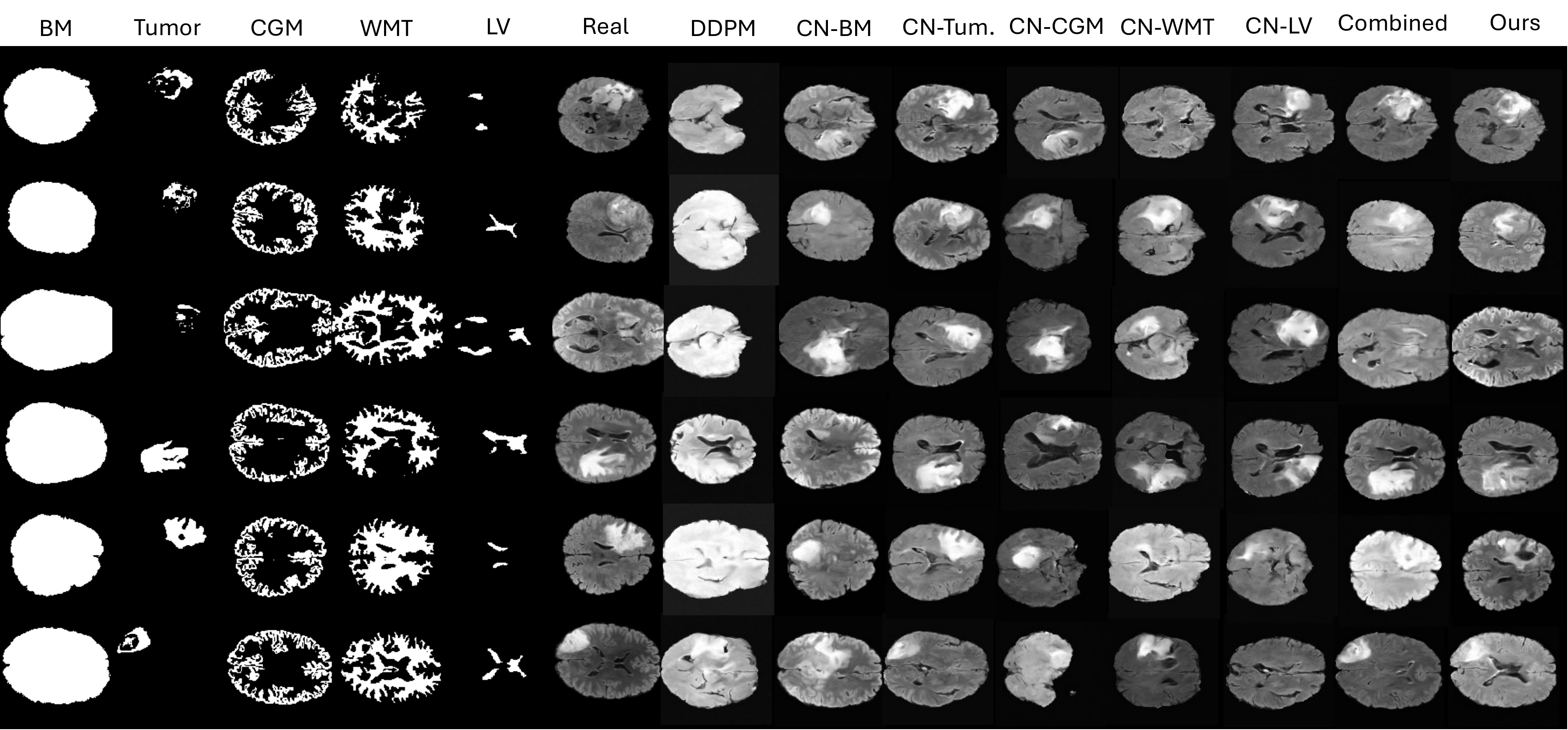}
    \caption{\textbf{Additional Baseline Comparisons.} Results from various baseline models, including DDPM and ControlNet trained with different controls: BM, tumor, CGM, WMT, and LV. We also present results using combined structures as controls for the ControlNet model, alongside our proposed method. Our approach produces anatomically accurate, high-quality images.}
    \label{fig:esupp3}
\end{figure*}

The supplementary presents the following materials:
\begin{itemize}
    \item Additional qualitative results (Figure \ref{fig:supp1}),
    \item Additional examples of brain anatomy structures and tumor segmentation masks (Figure \ref{fig:esupp2}), and
    \item Additional baseline comparisons (Figure \ref{fig:esupp3}).
\end{itemize}

See next page

%% file: main.bbl
\begin{thebibliography}{84}
\providecommand{\natexlab}[1]{#1}
\providecommand{\url}[1]{\texttt{#1}}
\expandafter\ifx\csname urlstyle\endcsname\relax
  \providecommand{\doi}[1]{doi: #1}\else
  \providecommand{\doi}{doi: \begingroup \urlstyle{rm}\Url}\fi

\bibitem[Alex et~al.(2017)Alex, KP, Chennamsetty, and Krishnamurthi]{alex2017generative}
Varghese Alex, Mohammed~Safwan KP, Sai~Saketh Chennamsetty, and Ganapathy Krishnamurthi.
\newblock Generative adversarial networks for brain lesion detection.
\newblock In \emph{Medical Imaging 2017: Image Processing}, pages 113--121. SPIE, 2017.

\bibitem[Alexanderson et~al.(2023)Alexanderson, Nagy, Beskow, and Henter]{alexanderson2023listen}
Simon Alexanderson, Rajmund Nagy, Jonas Beskow, and Gustav~Eje Henter.
\newblock Listen, denoise, action! audio-driven motion synthesis with diffusion models.
\newblock \emph{ACM Transactions on Graphics (TOG)}, 42\penalty0 (4):\penalty0 1--20, 2023.

\bibitem[Asaad et~al.(2022)Asaad, Ali, Majeed, and Rashid]{asaad2022persistent}
Aras Asaad, Dashti Ali, Taban Majeed, and Rasber Rashid.
\newblock Persistent homology for breast tumor classification using mammogram scans.
\newblock \emph{Mathematics}, 10\penalty0 (21):\penalty0 4039, 2022.

\bibitem[Baid et~al.(2021)Baid, Ghodasara, Mohan, Bilello, Calabrese, Colak, Farahani, Kalpathy-Cramer, Kitamura, Pati, et~al.]{baid2021rsna}
Ujjwal Baid, Satyam Ghodasara, Suyash Mohan, Michel Bilello, Evan Calabrese, Errol Colak, Keyvan Farahani, Jayashree Kalpathy-Cramer, Felipe~C Kitamura, Sarthak Pati, et~al.
\newblock The rsna-asnr-miccai brats 2021 benchmark on brain tumor segmentation and radiogenomic classification.
\newblock \emph{arXiv preprint arXiv:2107.02314}, 2021.

\bibitem[Banerjee et~al.(2020)Banerjee, Magee, Wang, Li, Huo, Jayakumar, Matho, Lin, Ram, Sivaprakasam, et~al.]{banerjee2020semantic}
Samik Banerjee, Lucas Magee, Dingkang Wang, Xu Li, Bing-Xing Huo, Jaikishan Jayakumar, Katherine Matho, Meng-Kuan Lin, Keerthi Ram, Mohanasankar Sivaprakasam, et~al.
\newblock Semantic segmentation of microscopic neuroanatomical data by combining topological priors with encoder--decoder deep networks.
\newblock \emph{Nature machine intelligence}, 2\penalty0 (10):\penalty0 585--594, 2020.

\bibitem[Batzies and Welker(2002)]{batzies2002discrete}
Ekkehard Batzies and Volkmar Welker.
\newblock Discrete morse theory for cellular resolutions.
\newblock 2002.

\bibitem[Behrendt et~al.(2025)Behrendt, Bhattacharya, Mieling, Maack, Kr{\"u}ger, Opfer, and Schlaefer]{behrendt2025guided}
Finn Behrendt, Debayan Bhattacharya, Robin Mieling, Lennart Maack, Julia Kr{\"u}ger, Roland Opfer, and Alexander Schlaefer.
\newblock Guided reconstruction with conditioned diffusion models for unsupervised anomaly detection in brain mris.
\newblock \emph{Computers in Biology and Medicine}, 186:\penalty0 109660, 2025.

\bibitem[Berger et~al.(2024)Berger, Lux, Stucki, B{\"u}rgin, Shit, Banaszak, Rueckert, Bauer, and Paetzold]{berger2024topologically}
Alexander~H Berger, Laurin Lux, Nico Stucki, Vincent B{\"u}rgin, Suprosanna Shit, Anna Banaszak, Daniel Rueckert, Ulrich Bauer, and Johannes~C Paetzold.
\newblock Topologically faithful multi-class segmentation in medical images.
\newblock In \emph{International Conference on Medical Image Computing and Computer-Assisted Intervention}, pages 721--731. Springer, 2024.

\bibitem[Bhattacharya and Prasanna(2024)]{bhattacharya2024gazediff}
Moinak Bhattacharya and Prateek Prasanna.
\newblock Gazediff: a radiologist visual attention guided diffusion model for zero-shot disease classification.
\newblock In \emph{Medical Imaging with Deep Learning}, 2024.

\bibitem[Bhattacharya et~al.(2024)Bhattacharya, Singh, Jain, and Prasanna]{bhattacharya2024radgazegen}
Moinak Bhattacharya, Gagandeep Singh, Shubham Jain, and Prateek Prasanna.
\newblock Radgazegen: Radiomics and gaze-guided medical image generation using diffusion models.
\newblock \emph{arXiv preprint arXiv:2410.00307}, 2024.

\bibitem[Billot et~al.(2023)Billot, Greve, Puonti, Thielscher, Van~Leemput, Fischl, Dalca, Iglesias, et~al.]{billot2023synthseg}
Benjamin Billot, Douglas~N Greve, Oula Puonti, Axel Thielscher, Koen Van~Leemput, Bruce Fischl, Adrian~V Dalca, Juan~Eugenio Iglesias, et~al.
\newblock Synthseg: Segmentation of brain mri scans of any contrast and resolution without retraining.
\newblock \emph{Medical image analysis}, 86:\penalty0 102789, 2023.

\bibitem[Bray et~al.(2024)Bray, Laversanne, Sung, Ferlay, Siegel, Soerjomataram, and Jemal]{bray2024global}
Freddie Bray, Mathieu Laversanne, Hyuna Sung, Jacques Ferlay, Rebecca~L Siegel, Isabelle Soerjomataram, and Ahmedin Jemal.
\newblock Global cancer statistics 2022: Globocan estimates of incidence and mortality worldwide for 36 cancers in 185 countries.
\newblock \emph{CA: a cancer journal for clinicians}, 74\penalty0 (3):\penalty0 229--263, 2024.

\bibitem[Cardoso et~al.(2022)Cardoso, Li, Brown, Ma, Kerfoot, Wang, Murrey, Myronenko, Zhao, Yang, et~al.]{cardoso2022monai}
M~Jorge Cardoso, Wenqi Li, Richard Brown, Nic Ma, Eric Kerfoot, Yiheng Wang, Benjamin Murrey, Andriy Myronenko, Can Zhao, Dong Yang, et~al.
\newblock Monai: An open-source framework for deep learning in healthcare.
\newblock \emph{arXiv preprint arXiv:2211.02701}, 2022.

\bibitem[Carlsson(2009)]{carlsson2009topology}
Gunnar Carlsson.
\newblock Topology and data.
\newblock \emph{Bulletin of the American Mathematical Society}, 46\penalty0 (2):\penalty0 255--308, 2009.

\bibitem[Chambon et~al.(2022)Chambon, Bluethgen, Delbrouck, Van~der Sluijs, Po{\l}acin, Chaves, Abraham, Purohit, Langlotz, and Chaudhari]{chambon2022roentgen}
Pierre Chambon, Christian Bluethgen, Jean-Benoit Delbrouck, Rogier Van~der Sluijs, Ma{\l}gorzata Po{\l}acin, Juan Manuel~Zambrano Chaves, Tanishq~Mathew Abraham, Shivanshu Purohit, Curtis~P Langlotz, and Akshay Chaudhari.
\newblock Roentgen: vision-language foundation model for chest x-ray generation.
\newblock \emph{arXiv preprint arXiv:2211.12737}, 2022.

\bibitem[Chen et~al.(2025)Chen, Tian, Wu, Feng, Lao, Zhang, Liao, and Wei]{chen2025joint}
Lixuan Chen, Xuanyu Tian, Jiangjie Wu, Ruimin Feng, Guoyan Lao, Yuyao Zhang, Hongen Liao, and Hongjiang Wei.
\newblock Joint coil sensitivity and motion correction in parallel mri with a self-calibrating score-based diffusion model.
\newblock \emph{Medical Image Analysis}, page 103502, 2025.

\bibitem[Chen et~al.(2023)Chen, Wang, and Shan]{chen2023berdiff}
Tao Chen, Chenhui Wang, and Hongming Shan.
\newblock Berdiff: Conditional bernoulli diffusion model for medical image segmentation.
\newblock In \emph{International conference on medical image computing and computer-assisted intervention}, pages 491--501. Springer, 2023.

\bibitem[Chu et~al.(2023)Chu, Zhou, Luo, Qiu, and Gao]{chu2023topology}
Yuetan Chu, Longxi Zhou, Gongning Luo, Zhaowen Qiu, and Xin Gao.
\newblock Topology-preserving computed tomography super-resolution based on dual-stream diffusion model.
\newblock In \emph{International Conference on Medical Image Computing and Computer-Assisted Intervention}, pages 260--270. Springer, 2023.

\bibitem[Dey et~al.(2019)Dey, Wang, and Wang]{dey2019road}
Tamal~K Dey, Jiayuan Wang, and Yusu Wang.
\newblock Road network reconstruction from satellite images with machine learning supported by topological methods.
\newblock In \emph{Proceedings of the 27th ACM SIGSPATIAL International Conference on Advances in Geographic Information Systems}, pages 520--523, 2019.

\bibitem[Dong et~al.(2024)Dong, Yuan, Hua, and Li]{dong2024diffusion}
Zhiwei Dong, Genji Yuan, Zhen Hua, and Jinjiang Li.
\newblock Diffusion model-based text-guided enhancement network for medical image segmentation.
\newblock \emph{Expert Systems with Applications}, 249:\penalty0 123549, 2024.

\bibitem[Edelsbrunner et~al.(2002)Edelsbrunner, Letscher, and Zomorodian]{edelsbrunner2002topological}
Edelsbrunner, Letscher, and Zomorodian.
\newblock Topological persistence and simplification.
\newblock \emph{Discrete \& computational geometry}, 28:\penalty0 511--533, 2002.

\bibitem[Edelsbrunner and Harer(2010)]{edelsbrunner2010computational}
Herbert Edelsbrunner and John Harer.
\newblock \emph{Computational topology: an introduction}.
\newblock American Mathematical Soc., 2010.

\bibitem[Goodfellow et~al.(2020)Goodfellow, Pouget-Abadie, Mirza, Xu, Warde-Farley, Ozair, Courville, and Bengio]{goodfellow2020generative}
Ian Goodfellow, Jean Pouget-Abadie, Mehdi Mirza, Bing Xu, David Warde-Farley, Sherjil Ozair, Aaron Courville, and Yoshua Bengio.
\newblock Generative adversarial networks.
\newblock \emph{Communications of the ACM}, 63\penalty0 (11):\penalty0 139--144, 2020.

\bibitem[Guo et~al.(2023)Guo, Yang, Ye, Lu, Peng, Huang, Xiang, and Ma]{guo2023accelerating}
Xutao Guo, Yanwu Yang, Chenfei Ye, Shang Lu, Bo Peng, Hua Huang, Yang Xiang, and Ting Ma.
\newblock Accelerating diffusion models via pre-segmentation diffusion sampling for medical image segmentation.
\newblock In \emph{2023 IEEE 20th International Symposium on Biomedical Imaging (ISBI)}, pages 1--5. IEEE, 2023.

\bibitem[Gupta et~al.(2022)Gupta, Hu, Kaan, Jin, Mpoy, Chung, Singh, Saltz, Kurc, Saltz, et~al.]{gupta2022learning}
Saumya Gupta, Xiaoling Hu, James Kaan, Michael Jin, Mutshipay Mpoy, Katherine Chung, Gagandeep Singh, Mary Saltz, Tahsin Kurc, Joel Saltz, et~al.
\newblock Learning topological interactions for multi-class medical image segmentation.
\newblock In \emph{European Conference on Computer Vision}, pages 701--718. Springer, 2022.

\bibitem[Gupta et~al.(2024)Gupta, Samaras, and Chen]{gupta2024topodiffusionnet}
Saumya Gupta, Dimitris Samaras, and Chao Chen.
\newblock Topodiffusionnet: A topology-aware diffusion model.
\newblock \emph{arXiv preprint arXiv:2410.16646}, 2024.

\bibitem[Ho et~al.(2020)Ho, Jain, and Abbeel]{ho2020denoising}
Jonathan Ho, Ajay Jain, and Pieter Abbeel.
\newblock Denoising diffusion probabilistic models.
\newblock \emph{Advances in neural information processing systems}, 33:\penalty0 6840--6851, 2020.

\bibitem[Hu et~al.(2024)Hu, Fei, Xu, Hou, Yang, Wang, Lei, Qian, and He]{hu2024topology}
Jiangbei Hu, Ben Fei, Baixin Xu, Fei Hou, Weidong Yang, Shengfa Wang, Na Lei, Chen Qian, and Ying He.
\newblock Topology-aware latent diffusion for 3d shape generation.
\newblock \emph{arXiv preprint arXiv:2401.17603}, 2024.

\bibitem[Hu et~al.(2021{\natexlab{a}})Hu, Lei, Wang, Wang, Feng, and Shen]{hu2021bidirectional}
Shengye Hu, Baiying Lei, Shuqiang Wang, Yong Wang, Zhiguang Feng, and Yanyan Shen.
\newblock Bidirectional mapping generative adversarial networks for brain mr to pet synthesis.
\newblock \emph{IEEE Transactions on Medical Imaging}, 41\penalty0 (1):\penalty0 145--157, 2021{\natexlab{a}}.

\bibitem[Hu et~al.(2019)Hu, Li, Samaras, and Chen]{hu2019topology}
Xiaoling Hu, Fuxin Li, Dimitris Samaras, and Chao Chen.
\newblock Topology-preserving deep image segmentation.
\newblock \emph{Advances in neural information processing systems}, 32, 2019.

\bibitem[Hu et~al.(2021{\natexlab{b}})Hu, Wang, Fuxin, Samaras, and Chen]{hu2021topology}
Xiaoling Hu, Yusu Wang, Li Fuxin, Dimitris Samaras, and Chao Chen.
\newblock Topology-aware segmentation using discrete morse theory.
\newblock \emph{arXiv preprint arXiv:2103.09992}, 2021{\natexlab{b}}.

\bibitem[Huang et~al.(2023)Huang, Chen, Liu, Shen, Zhao, and Zhou]{huang2023composer}
Lianghua Huang, Di Chen, Yu Liu, Yujun Shen, Deli Zhao, and Jingren Zhou.
\newblock Composer: Creative and controllable image synthesis with composable conditions.
\newblock \emph{arXiv preprint arXiv:2302.09778}, 2023.

\bibitem[Kachouie(2008)]{kachouie2008anisotropic}
Nezamoddin~N Kachouie.
\newblock Anisotropic diffusion for medical image enhancement.
\newblock \emph{Int. J. Image Process}, 4\penalty0 (4):\penalty0 436--443, 2008.

\bibitem[Kazerouni et~al.(2023)Kazerouni, Aghdam, Heidari, Azad, Fayyaz, Hacihaliloglu, and Merhof]{kazerouni2023diffusion}
Amirhossein Kazerouni, Ehsan~Khodapanah Aghdam, Moein Heidari, Reza Azad, Mohsen Fayyaz, Ilker Hacihaliloglu, and Dorit Merhof.
\newblock Diffusion models in medical imaging: A comprehensive survey.
\newblock \emph{Medical image analysis}, 88:\penalty0 102846, 2023.

\bibitem[Khader et~al.(2023)Khader, M{\"u}ller-Franzes, Tayebi~Arasteh, Han, Haarburger, Schulze-Hagen, Schad, Engelhardt, Bae{\ss}ler, Foersch, et~al.]{khader2023denoising}
Firas Khader, Gustav M{\"u}ller-Franzes, Soroosh Tayebi~Arasteh, Tianyu Han, Christoph Haarburger, Maximilian Schulze-Hagen, Philipp Schad, Sandy Engelhardt, Bettina Bae{\ss}ler, Sebastian Foersch, et~al.
\newblock Denoising diffusion probabilistic models for 3d medical image generation.
\newblock \emph{Scientific Reports}, 13\penalty0 (1):\penalty0 7303, 2023.

\bibitem[Kim and Ye(2022)]{kim2022diffusion}
Boah Kim and Jong~Chul Ye.
\newblock Diffusion deformable model for 4d temporal medical image generation.
\newblock In \emph{International Conference on Medical Image Computing and Computer-Assisted Intervention}, pages 539--548. Springer, 2022.

\bibitem[Konz et~al.(2024)Konz, Chen, Dong, and Mazurowski]{konz2024anatomically}
Nicholas Konz, Yuwen Chen, Haoyu Dong, and Maciej~A Mazurowski.
\newblock Anatomically-controllable medical image generation with segmentation-guided diffusion models.
\newblock In \emph{International Conference on Medical Image Computing and Computer-Assisted Intervention}, pages 88--98. Springer, 2024.

\bibitem[Kwon et~al.(2019)Kwon, Han, and Kim]{kwon2019generation}
Gihyun Kwon, Chihye Han, and Dae-shik Kim.
\newblock Generation of 3d brain mri using auto-encoding generative adversarial networks.
\newblock In \emph{International Conference on Medical Image Computing and Computer-Assisted Intervention}, pages 118--126. Springer, 2019.

\bibitem[Lee et~al.(2024)Lee, Choi, Lim, Kim, and Shim]{lee2024scribble}
Seonho Lee, Jiho Choi, Seohyun Lim, Jiwook Kim, and Hyunjung Shim.
\newblock Scribble-guided diffusion for training-free text-to-image generation.
\newblock \emph{arXiv preprint arXiv:2409.08026}, 2024.

\bibitem[Li et~al.(2023)Li, Liu, Wu, Mu, Yang, Gao, Li, and Lee]{li2023gligen}
Yuheng Li, Haotian Liu, Qingyang Wu, Fangzhou Mu, Jianwei Yang, Jianfeng Gao, Chunyuan Li, and Yong~Jae Lee.
\newblock Gligen: Open-set grounded text-to-image generation.
\newblock In \emph{Proceedings of the IEEE/CVF conference on computer vision and pattern recognition}, pages 22511--22521, 2023.

\bibitem[Liu and Huang(2023)]{liu2023esdiff}
Fengting Liu and Wenhui Huang.
\newblock Esdiff: a joint model for low-quality retinal image enhancement and vessel segmentation using a diffusion model.
\newblock \emph{Biomedical Optics Express}, 14\penalty0 (12):\penalty0 6563--6578, 2023.

\bibitem[Ma et~al.(2023)Ma, Zhu, You, and Wang]{ma2023pre}
Jun Ma, Yuanzhi Zhu, Chenyu You, and Bo Wang.
\newblock Pre-trained diffusion models for plug-and-play medical image enhancement.
\newblock In \emph{International Conference on Medical Image Computing and Computer-Assisted Intervention}, pages 3--13. Springer, 2023.

\bibitem[Ma et~al.(2025)Ma, Jian, and Chen]{ma2025diffusion}
Ji Ma, Guojun Jian, and Jinjin Chen.
\newblock Diffusion model-based mri super-resolution synthesis.
\newblock \emph{International Journal of Imaging Systems and Technology}, 35\penalty0 (2):\penalty0 e70021, 2025.

\bibitem[Ma et~al.(2024)Ma, He, Cun, Wang, Chen, Li, and Chen]{ma2024follow}
Yue Ma, Yingqing He, Xiaodong Cun, Xintao Wang, Siran Chen, Xiu Li, and Qifeng Chen.
\newblock Follow your pose: Pose-guided text-to-video generation using pose-free videos.
\newblock In \emph{Proceedings of the AAAI Conference on Artificial Intelligence}, pages 4117--4125, 2024.

\bibitem[Moawad et~al.(2024)Moawad, Janas, Baid, Ramakrishnan, Saluja, Ashraf, Jekel, Amiruddin, Adewole, Albrecht, et~al.]{moawad2024brain}
Ahmed~W Moawad, Anastasia Janas, Ujjwal Baid, Divya Ramakrishnan, Rachit Saluja, Nader Ashraf, Leon Jekel, Raisa Amiruddin, Maruf Adewole, Jake Albrecht, et~al.
\newblock The brain tumor segmentation-metastases (brats-mets) challenge 2023: Brain metastasis segmentation on pre-treatment mri.
\newblock \emph{ArXiv}, pages arXiv--2306, 2024.

\bibitem[Mou et~al.(2024)Mou, Wang, Xie, Wu, Zhang, Qi, and Shan]{mou2024t2i}
Chong Mou, Xintao Wang, Liangbin Xie, Yanze Wu, Jian Zhang, Zhongang Qi, and Ying Shan.
\newblock T2i-adapter: Learning adapters to dig out more controllable ability for text-to-image diffusion models.
\newblock In \emph{Proceedings of the AAAI Conference on Artificial Intelligence}, pages 4296--4304, 2024.

\bibitem[Panaretos and Zemel(2019)]{panaretos2019statistical}
Victor~M Panaretos and Yoav Zemel.
\newblock Statistical aspects of wasserstein distances.
\newblock \emph{Annual review of statistics and its application}, 6\penalty0 (1):\penalty0 405--431, 2019.

\bibitem[Park et~al.()Park, Lee, Song, Wu, and Kim]{parktopology}
Joonhyuk Park, Donghyun Lee, Yujee Song, Guorong Wu, and Won~Hwa Kim.
\newblock Topology-aware graph diffusion model with persistent homology.

\bibitem[Park et~al.(2019)Park, Liu, Wang, and Zhu]{park2019semantic}
Taesung Park, Ming-Yu Liu, Ting-Chun Wang, and Jun-Yan Zhu.
\newblock Semantic image synthesis with spatially-adaptive normalization.
\newblock In \emph{Proceedings of the IEEE/CVF conference on computer vision and pattern recognition}, pages 2337--2346, 2019.

\bibitem[Paszke et~al.(2019)Paszke, Gross, Massa, Lerer, Bradbury, Chanan, Killeen, Lin, Gimelshein, Antiga, et~al.]{paszke2019pytorch}
Adam Paszke, Sam Gross, Francisco Massa, Adam Lerer, James Bradbury, Gregory Chanan, Trevor Killeen, Zeming Lin, Natalia Gimelshein, Luca Antiga, et~al.
\newblock Pytorch: An imperative style, high-performance deep learning library.
\newblock \emph{Advances in neural information processing systems}, 32, 2019.

\bibitem[Peng et~al.(2024)Peng, Wang, Sonka, and Chen]{peng2024phg}
Yaopeng Peng, Hongxiao Wang, Milan Sonka, and Danny~Z Chen.
\newblock Phg-net: Persistent homology guided medical image classification.
\newblock In \emph{Proceedings of the IEEE/CVF Winter Conference on Applications of Computer Vision}, pages 7583--7592, 2024.

\bibitem[Pinaya et~al.(2022)Pinaya, Tudosiu, Dafflon, Da~Costa, Fernandez, Nachev, Ourselin, and Cardoso]{pinaya2022brain}
Walter~HL Pinaya, Petru-Daniel Tudosiu, Jessica Dafflon, Pedro~F Da~Costa, Virginia Fernandez, Parashkev Nachev, Sebastien Ourselin, and M~Jorge Cardoso.
\newblock Brain imaging generation with latent diffusion models.
\newblock In \emph{MICCAI Workshop on Deep Generative Models}, pages 117--126. Springer, 2022.

\bibitem[Pinaya et~al.(2023)Pinaya, Graham, Kerfoot, Tudosiu, Dafflon, Fernandez, Sanchez, Wolleb, Da~Costa, Patel, et~al.]{pinaya2023generative}
Walter~HL Pinaya, Mark~S Graham, Eric Kerfoot, Petru-Daniel Tudosiu, Jessica Dafflon, Virginia Fernandez, Pedro Sanchez, Julia Wolleb, Pedro~F Da~Costa, Ashay Patel, et~al.
\newblock Generative ai for medical imaging: extending the monai framework.
\newblock \emph{arXiv preprint arXiv:2307.15208}, 2023.

\bibitem[Qin et~al.(2023)Qin, Zhang, Yu, Feng, Yang, Zhou, Wang, Niebles, Xiong, Savarese, et~al.]{qin2023unicontrol}
Can Qin, Shu Zhang, Ning Yu, Yihao Feng, Xinyi Yang, Yingbo Zhou, Huan Wang, Juan~Carlos Niebles, Caiming Xiong, Silvio Savarese, et~al.
\newblock Unicontrol: A unified diffusion model for controllable visual generation in the wild.
\newblock \emph{arXiv preprint arXiv:2305.11147}, 2023.

\bibitem[Qin et~al.(2025)Qin, Xu, Zhang, Xiong, Yuan, and He]{qin2025btsegdiff}
Jiacheng Qin, Dan Xu, Hao Zhang, Zhaoyu Xiong, Yejing Yuan, and Kangjian He.
\newblock Btsegdiff: Brain tumor segmentation based on multimodal mri dynamically guided diffusion probability model.
\newblock \emph{Computers in Biology and Medicine}, 186:\penalty0 109694, 2025.

\bibitem[Radford et~al.(2021)Radford, Kim, Hallacy, Ramesh, Goh, Agarwal, Sastry, Askell, Mishkin, Clark, et~al.]{radford2021learning}
Alec Radford, Jong~Wook Kim, Chris Hallacy, Aditya Ramesh, Gabriel Goh, Sandhini Agarwal, Girish Sastry, Amanda Askell, Pamela Mishkin, Jack Clark, et~al.
\newblock Learning transferable visual models from natural language supervision.
\newblock In \emph{International conference on machine learning}, pages 8748--8763. PMLR, 2021.

\bibitem[Rahman et~al.(2023)Rahman, Valanarasu, Hacihaliloglu, and Patel]{rahman2023ambiguous}
Aimon Rahman, Jeya Maria~Jose Valanarasu, Ilker Hacihaliloglu, and Vishal~M Patel.
\newblock Ambiguous medical image segmentation using diffusion models.
\newblock In \emph{Proceedings of the IEEE/CVF conference on computer vision and pattern recognition}, pages 11536--11546, 2023.

\bibitem[Rombach et~al.(2022)Rombach, Blattmann, Lorenz, Esser, and Ommer]{rombach2022high}
Robin Rombach, Andreas Blattmann, Dominik Lorenz, Patrick Esser, and Bj{\"o}rn Ommer.
\newblock High-resolution image synthesis with latent diffusion models.
\newblock In \emph{Proceedings of the IEEE/CVF conference on computer vision and pattern recognition}, pages 10684--10695, 2022.

\bibitem[Ruiz et~al.(2023)Ruiz, Li, Jampani, Pritch, Rubinstein, and Aberman]{ruiz2023dreambooth}
Nataniel Ruiz, Yuanzhen Li, Varun Jampani, Yael Pritch, Michael Rubinstein, and Kfir Aberman.
\newblock Dreambooth: Fine tuning text-to-image diffusion models for subject-driven generation.
\newblock In \emph{Proceedings of the IEEE/CVF conference on computer vision and pattern recognition}, pages 22500--22510, 2023.

\bibitem[Saharia et~al.(2022)Saharia, Chan, Saxena, Li, Whang, Denton, Ghasemipour, Gontijo~Lopes, Karagol~Ayan, Salimans, et~al.]{saharia2022photorealistic}
Chitwan Saharia, William Chan, Saurabh Saxena, Lala Li, Jay Whang, Emily~L Denton, Kamyar Ghasemipour, Raphael Gontijo~Lopes, Burcu Karagol~Ayan, Tim Salimans, et~al.
\newblock Photorealistic text-to-image diffusion models with deep language understanding.
\newblock \emph{Advances in neural information processing systems}, 35:\penalty0 36479--36494, 2022.

\bibitem[Selim et~al.(2023)Selim, Zhang, Fathi, Brooks, Wang, Yu, and Chen]{selim2023latent}
Md Selim, Jie Zhang, Faraneh Fathi, Michael~A Brooks, Ge Wang, Guoqiang Yu, and Jin Chen.
\newblock Latent diffusion model for medical image standardization and enhancement.
\newblock \emph{arXiv preprint arXiv:2310.05237}, 2023.

\bibitem[Sharma et~al.(2024)Sharma, Kumar, Jha, Bhuyan, Das, and Bagci]{sharma2024controlpolypnet}
Vanshali Sharma, Abhishek Kumar, Debesh Jha, Manas~Kamal Bhuyan, Pradip~K Das, and Ulas Bagci.
\newblock Controlpolypnet: towards controlled colon polyp synthesis for improved polyp segmentation.
\newblock In \emph{Proceedings of the IEEE/CVF Conference on Computer Vision and Pattern Recognition}, pages 2325--2334, 2024.

\bibitem[Shen et~al.(2023{\natexlab{a}})Shen, Ye, Zhang, Wang, Han, and Yang]{shen2023advancing}
Fei Shen, Hu Ye, Jun Zhang, Cong Wang, Xiao Han, and Wei Yang.
\newblock Advancing pose-guided image synthesis with progressive conditional diffusion models.
\newblock \emph{arXiv preprint arXiv:2310.06313}, 2023{\natexlab{a}}.

\bibitem[Shen et~al.(2023{\natexlab{b}})Shen, Zhao, Meng, Li, Zhu, Zhou, and Lu]{shen2023difftalk}
Shuai Shen, Wenliang Zhao, Zibin Meng, Wanhua Li, Zheng Zhu, Jie Zhou, and Jiwen Lu.
\newblock Difftalk: Crafting diffusion models for generalized audio-driven portraits animation.
\newblock In \emph{Proceedings of the IEEE/CVF Conference on Computer Vision and Pattern Recognition}, pages 1982--1991, 2023{\natexlab{b}}.

\bibitem[Shi et~al.(2024)Shi, Hu, Yang, Gao, Liu, and Ma]{shi2024centerline}
Pengcheng Shi, Jiesi Hu, Yanwu Yang, Zilve Gao, Wei Liu, and Ting Ma.
\newblock Centerline boundary dice loss for vascular segmentation.
\newblock In \emph{International Conference on Medical Image Computing and Computer-Assisted Intervention}, pages 46--56. Springer, 2024.

\bibitem[Shit et~al.(2021)Shit, Paetzold, Sekuboyina, Ezhov, Unger, Zhylka, Pluim, Bauer, and Menze]{shit2021cldice}
Suprosanna Shit, Johannes~C Paetzold, Anjany Sekuboyina, Ivan Ezhov, Alexander Unger, Andrey Zhylka, Josien~PW Pluim, Ulrich Bauer, and Bjoern~H Menze.
\newblock cldice-a novel topology-preserving loss function for tubular structure segmentation.
\newblock In \emph{Proceedings of the IEEE/CVF conference on computer vision and pattern recognition}, pages 16560--16569, 2021.

\bibitem[Singh et~al.(2014)Singh, Couture, Marron, Perou, and Niethammer]{singh2014topological}
Nikhil Singh, Heather~D Couture, JS Marron, Charles Perou, and Marc Niethammer.
\newblock Topological descriptors of histology images.
\newblock In \emph{Machine Learning in Medical Imaging: 5th International Workshop, MLMI 2014, Held in Conjunction with MICCAI 2014, Boston, MA, USA, September 14, 2014. Proceedings 5}, pages 231--239. Springer, 2014.

\bibitem[Song et~al.(2024)Song, Zhan, Chen, and Shi]{song2024topo}
Chen Song, Wenkang Zhan, Yuzhou Chen, and Xinghua Shi.
\newblock Topo-diffusion: Topological diffusion model for image and point cloud generation.
\newblock 2024.

\bibitem[Song et~al.(2020)Song, Sohl-Dickstein, Kingma, Kumar, Ermon, and Poole]{song2020score}
Yang Song, Jascha Sohl-Dickstein, Diederik~P Kingma, Abhishek Kumar, Stefano Ermon, and Ben Poole.
\newblock Score-based generative modeling through stochastic differential equations.
\newblock \emph{arXiv preprint arXiv:2011.13456}, 2020.

\bibitem[Sun et~al.(2022)Sun, Han, Kong, Tang, Yan, and Xie]{sun2022topology}
Shanlin Sun, Kun Han, Deying Kong, Hao Tang, Xiangyi Yan, and Xiaohui Xie.
\newblock Topology-preserving shape reconstruction and registration via neural diffeomorphic flow.
\newblock In \emph{Proceedings of the IEEE/CVF Conference on Computer Vision and Pattern Recognition}, pages 20845--20855, 2022.

\bibitem[Tseng et~al.(2023)Tseng, Li, Kim, Alsisan, Huang, and Kopf]{tseng2023consistent}
Hung-Yu Tseng, Qinbo Li, Changil Kim, Suhib Alsisan, Jia-Bin Huang, and Johannes Kopf.
\newblock Consistent view synthesis with pose-guided diffusion models.
\newblock In \emph{Proceedings of the IEEE/CVF Conference on Computer Vision and Pattern Recognition}, pages 16773--16783, 2023.

\bibitem[Wang et~al.(2025)Wang, Zou, Sakla, Partyka, Rawal, Singh, Zhao, Ling, Huang, Prasanna, et~al.]{wang2025topotxr}
Fan Wang, Zhilin Zou, Nicole Sakla, Luke Partyka, Nil Rawal, Gagandeep Singh, Wei Zhao, Haibin Ling, Chuan Huang, Prateek Prasanna, et~al.
\newblock Topotxr: A topology-guided deep convolutional network for breast parenchyma learning on dce-mris.
\newblock \emph{Medical Image Analysis}, 99:\penalty0 103373, 2025.

\bibitem[Wang et~al.(2022)Wang, Xian, and Vakanski]{wang2022ta}
Haotian Wang, Min Xian, and Aleksandar Vakanski.
\newblock Ta-net: Topology-aware network for gland segmentation.
\newblock In \emph{Proceedings of the IEEE/CVF winter conference on applications of computer vision}, pages 1556--1564, 2022.

\bibitem[Wolleb et~al.(2022)Wolleb, Bieder, Sandk{\"u}hler, and Cattin]{wolleb2022diffusion}
Julia Wolleb, Florentin Bieder, Robin Sandk{\"u}hler, and Philippe~C Cattin.
\newblock Diffusion models for medical anomaly detection.
\newblock In \emph{International Conference on Medical image computing and computer-assisted intervention}, pages 35--45. Springer, 2022.

\bibitem[Wu et~al.(2024)Wu, Fu, Fang, Zhang, Yang, Xiong, Liu, and Xu]{wu2024medsegdiff}
Junde Wu, Rao Fu, Huihui Fang, Yu Zhang, Yehui Yang, Haoyi Xiong, Huiying Liu, and Yanwu Xu.
\newblock Medsegdiff: Medical image segmentation with diffusion probabilistic model.
\newblock In \emph{Medical Imaging with Deep Learning}, pages 1623--1639. PMLR, 2024.

\bibitem[Xu et~al.(2024)Xu, Gupta, Hu, Li, Abousamra, Samaras, Prasanna, and Chen]{xu2024topocellgen}
Meilong Xu, Saumya Gupta, Xiaoling Hu, Chen Li, Shahira Abousamra, Dimitris Samaras, Prateek Prasanna, and Chao Chen.
\newblock Topocellgen: Generating histopathology cell topology with a diffusion model.
\newblock \emph{arXiv preprint arXiv:2412.06011}, 2024.

\bibitem[Yadav et~al.(2023)Yadav, Ahmed, Daescu, Gedik, and Coskunuzer]{yadav2023histopathological}
Ankur Yadav, Faisal Ahmed, Ovidiu Daescu, Reyhan Gedik, and Baris Coskunuzer.
\newblock Histopathological cancer detection with topological signatures.
\newblock In \emph{2023 IEEE International Conference on Bioinformatics and Biomedicine (BIBM)}, pages 1610--1619. IEEE, 2023.

\bibitem[Yang et~al.(2024)Yang, Musio, Ma, Juchler, Paetzold, Al-Maskari, H{\"o}her, Li, Hamamci, Sekuboyina, et~al.]{yang2024benchmarking}
Kaiyuan Yang, Fabio Musio, Yihui Ma, Norman Juchler, Johannes~C Paetzold, Rami Al-Maskari, Luciano H{\"o}her, Hongwei~Bran Li, Ibrahim~Ethem Hamamci, Anjany Sekuboyina, et~al.
\newblock Benchmarking the cow with the topcow challenge: Topology-aware anatomical segmentation of the circle of willis for cta and mra.
\newblock \emph{ArXiv}, pages arXiv--2312, 2024.

\bibitem[Yoon et~al.(2023)Yoon, Zhang, Suk, Guo, and Li]{yoon2023sadm}
Jee~Seok Yoon, Chenghao Zhang, Heung-Il Suk, Jia Guo, and Xiaoxiao Li.
\newblock Sadm: Sequence-aware diffusion model for longitudinal medical image generation.
\newblock In \emph{International Conference on Information Processing in Medical Imaging}, pages 388--400. Springer, 2023.

\bibitem[Zhang and Shi(2024)]{zhang2024anatomy}
Jianwei Zhang and Yonggang Shi.
\newblock Anatomy-guided surface diffusion model for alzheimer's disease normative modeling.
\newblock \emph{arXiv preprint arXiv:2403.04531}, 2024.

\bibitem[Zhang et~al.(2023)Zhang, Rao, and Agrawala]{zhang2023adding}
Lvmin Zhang, Anyi Rao, and Maneesh Agrawala.
\newblock Adding conditional control to text-to-image diffusion models.
\newblock In \emph{Proceedings of the IEEE/CVF international conference on computer vision}, pages 3836--3847, 2023.

\bibitem[Zhao et~al.(2023{\natexlab{a}})Zhao, Chen, Chen, Bao, Hao, Yuan, and Wong]{zhao2023uni}
Shihao Zhao, Dongdong Chen, Yen-Chun Chen, Jianmin Bao, Shaozhe Hao, Lu Yuan, and Kwan-Yee~K Wong.
\newblock Uni-controlnet: All-in-one control to text-to-image diffusion models.
\newblock \emph{Advances in Neural Information Processing Systems}, 36:\penalty0 11127--11150, 2023{\natexlab{a}}.

\bibitem[Zhao et~al.(2023{\natexlab{b}})Zhao, Chen, Fu, Yang, Ma, Zhu, Wang, Jiao, Jin, Xiao, et~al.]{zhao2023single}
Shen Zhao, De-Pin Chen, Tong Fu, Jing-Cheng Yang, Ding Ma, Xiu-Zhi Zhu, Xiang-Xue Wang, Yi-Ping Jiao, Xi Jin, Yi Xiao, et~al.
\newblock Single-cell morphological and topological atlas reveals the ecosystem diversity of human breast cancer.
\newblock \emph{Nature Communications}, 14\penalty0 (1):\penalty0 6796, 2023{\natexlab{b}}.

\bibitem[Zhu et~al.(2024)Zhu, Yu, Zhao, Liu, Ye, Chen, Zhang, Wei, and Liang]{zhu2024controltraj}
Yuanshao Zhu, James~Jianqiao Yu, Xiangyu Zhao, Qidong Liu, Yongchao Ye, Wei Chen, Zijian Zhang, Xuetao Wei, and Yuxuan Liang.
\newblock Controltraj: Controllable trajectory generation with topology-constrained diffusion model.
\newblock In \emph{Proceedings of the 30th ACM SIGKDD Conference on Knowledge Discovery and Data Mining}, pages 4676--4687, 2024.

\end{thebibliography}
